\documentclass[12pt]{JHEP3}

\usepackage{mathrsfs}
\usepackage{amsmath,amssymb}
\usepackage{epsfig}
\input epsf
\preprint{ {\tt hep-th/0502240}\\ {\tt ITP-05/09} \\ {\tt
SPIN-05/07}}
\def\x'{\mathaccent 19 x}
\def\y'{\mathaccent 19 y}
\def\n'{\mathaccent 19 n}
\def\u'{\mathaccent 19 u}

\def\X'{\mathaccent 19 X}
\def\Y'{\mathaccent 19 Y}
\def\Z'{\mathaccent 19 Z}

\def\et'{\mathaccent 19 \eta}
\def\th'{\mathaccent 19 \theta}
\def\lam'{\mathaccent 19 \lambda}
\def\varet'{\mathaccent 19 \vartheta}
\def\rh'{\mathaccent 19 \rho}
\def\ph'{\mathaccent 19 \phi}
\def\xb'{\mathaccent 19 {\bar{x}}}

\def\det{\hbox{det}}
\def\be{\begin{equation}}
\def\ee{\end{equation}}

\newcommand{\bea}{\begin{eqnarray}}
\newcommand{\eea}{\end{eqnarray}}

\def\a {\alpha}
\def\b {\beta}
\def\s {\sigma}
\def\pa {\partial}
\def\P {\mathscr{P}}

\def \gg{{\rm g}}

\def \la{\label}
\newcommand{\rf}[1]{(\ref{#1})}

\newcommand{\alg}[1]{\mathfrak{#1}}
\newcommand{\su}{\alg{su}}
\newcommand{\psu}{\alg{psu}}
\newcommand{\un}{\alg{u}}

\newcommand{\AdS}{{\rm  AdS}_5\times {\rm S}^5}

\def\L{\mathscr L}
\def \l {\lambda}
\def \JI {{\rm I}}
\title{On Integrability of Classical \\SuperStrings in
$\AdS$}

\author{L. F. Alday$^{a}$, G. Arutyunov$^{a}$ and A. A.
Tseytlin$^{b,}$\footnote{Also at Imperial College London
 and  Lebedev  Institute, Moscow
 }
\footnote{L.F.Alday@phys.uu.nl, G.Arutyunov@phys.uu.nl, tseytlin@mps.ohio-state.edu}\\
$^{a}$ {\it Institute for Theoretical Physics and Spinoza Institute, \\
~~Utrecht University, Netherlands}\\
$^{b}$ {\it Department of Physics,
The Ohio State University,\\
Columbus, OH 43210-1106, USA}}
 \abstract{We explore  integrability properties of
 superstring equations  of motion  in $\AdS$.
 We impose light-cone kappa-symmetry and reparametrization gauges
 and  construct  a
Lax representation for the corresponding  Hamiltonian dynamics on
subspace of physical superstring degrees of freedom. We present
some explicit results for the corresponding conserved  charges by
consistently reducing the dynamics to ${\rm AdS}_3 \times {\rm
S}^3$ and ${\rm AdS}_3 \times {\rm S}^1$ subsectors containing
both bosonic and fermionic fields.
 }

\begin{document}

\newpage

\def \adss {${\rm AdS}^5 \times {\rm S}^5$\ }
\def \ci{\cite}
\def \foot {\footnote}
\def \N {{\cal N}}
\def \l {\lambda}
\def \dD {{\rm D}}

\renewcommand{\thefootnote}{\arabic{footnote}}
\setcounter{footnote}{0}
\section{Introduction}
%%%%%%%%%%%%%%%%%%%%%%%%%%%%%%%%%%%%%%%%

The AdS/CFT duality  \ci{ads}
 between the $\N=4$ SYM theory
and the  $\AdS$  string theory implies
 various  relations between their respective properties.
One property that  attracted much attention  recently
is integrability. Both perturbative ($\l \to 0$)
planar  gauge theory  and the
classical ($\l \to \infty$) string theory on a 2-sphere
indicate the presence of integrability, suggesting that
it is a feature  of the theory at any finite value of `t Hooft
coupling $\l$  (proportional to the square of
string tension).

The string theory in   $\AdS$ is  defined by a fermionic
Green-Schwarz \ci{GS}  extension of the  bosonic coset  sigma
model \ci{MT}. The latter  is integrable  as a classical 2d field
theory in the sense of \ci{lu}. It is then  natural to expect
(given that
 the local kappa symmetry and global supersymmetry
   ``glue'' together the  bosonic and fermionic string coordinates,
   and also that the classical conformal symmetry of the string
   action should survive  quantum corrections thanks to fermionic
   contributions  \ci{MT})
   that the integrability  should play a prominent role
   in the full  quantum  world-sheet  theory
   defined on a 2-sphere.\foot{Potential importance
   of  integrability in
    $\AdS$ string theory was mentioned   in
    \cite{Metsaev:2000yu} and was also
 emphasized  in  \cite{Mandal:2002fs}.}

 In  \cite{Bena}
  it was explicitly verified
     that integrability  should be   present in the classical superstring theory
      by constructing the corresponding Lax pair (see also  \cite{Polyakov:2004br}
       for  related
       observations).
 The main issue is  how to extend this  to quantum theory.
 In contrast  to the  purely-bosonic
  coset cases where  integrability  does not actually survive at
  the quantum level (apart from the  case  of the principal chiral model),
here, due to  the quantum conformal symmetry,
   most of the relations implied by integrability
 should  indeed carry over to the quantum theory case
  ({\it i.e.} there should  be no non-trivial modification of the
algebra of conserved currents, etc.).\foot{See
 \cite{berk,berk2} for  a  discussion of this
 in the pure spinor approach.}

The  survival of integrability at  the
quantum string  level  is,  of course,
 strongly suggested  via the AdS/CFT by its
presence in the perturbative gauge theory.
%%%%%%%%%%%%%%%%%%%%%%%%%%%%%%%%%%%%%%%%%%%%%%%%%%%%%%%%%%%%%%%%%%%%%
Integrability on  perturbative gauge theory side (observed  already
  in a particular sector of QCD  at one loop \ci{qcd})
 in the  $\N=4$  SYM theory becomes  a  feature of the
 full dilatation operator \ci{mz} and
should  be present  to all loop orders  \ci{bks}
(see \cite{beis} for a review).
 One may conjecture  that it
survives  at any finite  value of the  't Hooft coupling and
 thus should translate into
the integrability  of $\AdS$  string theory.

%%%%%%%%%%%%%%%%%%%%%%%%%%%%%%%

To try to establish the matching of the two  integrable structures
one should note that the duality relates only physical, {\it i.e.}
gauge-invariant, quantities on the two sides (e.g.,  the SYM
theory does not know about gauge-dependent properties of string
theory and vice versa).\footnote{ This is illustrated, in
particular, on the example of   matching the coherent-state
Landau-Lifshitz model  for semiclassical spin chain states  to the
``fast-string'' limit  of the superstring action  (see \cite{kru}
for bosonic cases and  \cite{mik3} for cases including fermionic
degrees of  freedom).} One would like, therefore,
 to
exhibit the integrable structure of the $\AdS$ string theory  in a
physical gauge, where  quantization and, eventually, relation  to
gauge theory may become more explicit.

A natural physical  gauge choice is the  light-cone  $\kappa$-symmetry
 gauge   suggested
in \cite{Metsaev:2000yf}. Supplemented by  the  light-cone bosonic
gauge \cite{Metsaev:2000yu} adapted to Poincar\'e coordinates it
leads to  a very explicit form of the string dynamics, described
by a string action for 8+8 physical degrees of freedom which is
only of  quartic order in fermions. The phase-space approach of
\cite{Metsaev:2000yu} seems a natural starting point for
quantizing the $\AdS$ superstring.

The problem we are going to address  in this work is how to
construct explicitly the Lax representation for the classical
Hamiltonian $\AdS$ superstring equations in the light-cone gauge
of \cite{Metsaev:2000yu}. Due to the well-known difficulties with
the covariant Hamiltonian treatment of the $\kappa$-symmetric
string this question becomes particularly important for
understanding the integrable structure of  {\it quantum}
superstrings on $\AdS$. Indeed, having an explicit Lax
representation based on the Lax connection $\L$, which involves
only physical degrees of freedom, one can unambiguously determine
the Poisson brackets of the matrix elements of $\L$ and hopefully
encode them into the form of the classical $r$-matrix. In many
known examples the classical $r$-matrix structure is very helpful
to find the corresponding quantum theory \cite{FTa}.

The basic tool we will use in order to obtain the Lax
representation for the gauge-fixed Hamiltonian is the covariant
Lax connection for superstrings on $\AdS$ found in \cite{Bena}. We
will show that this connection admits a reduction to the {\it
physical subspace} determined by solutions of the (bosonic and
fermionic) gauge conditions and constraints. We realize the
connection explicitly in a ``minimal way" in terms of $8\times 8$
matrices from the superalgebra $\su(2,2|4)$. This realization
enables us to further investigate some spectral properties of the
associated monodromy matrix.

%%%%%%%%%%%%%%%%%%%%%%%%%%%%%%%%%%
Let us mention also that  related aspects of integrability of
$\AdS$ string theory  and its gauge theory counterpart
 were recently discussed, e.g.,  in \ci{afrt}-\ci{Das}.
%\ci{afrt,arst,Ka,Arutyunov:2004yx,ald,swan,Dol,Das}.
  %%%%%%%%%%%%%%%%%%%%%%%%%%%%%%%%%%%%%%%%%%%%%%%%%%%%%%%%%%%%%%%

\bigskip

The paper is organized as follows. In section 2 we shall review
the structure of the covariant $\AdS$ superstring equations of
motion \ci{MT,Roiban:2000yy} interpreted in terms of
 currents of the $\frac{{\rm PSU}(2,2|4)}{{\rm
SO}(4,1)\times {\rm SO}(5)}$ supercoset and identify the
corresponding Lax connection as in  \cite{Bena}. We will make some
general comments on the  form of the Lax connection in the
$8\times 8$ matrix $\su(2,2|4)$  representation and on its
asymptotic expansion in the spectral parameter.

In section 3 we will recall the form of the light-cone gauge fixed
action of \cite{Metsaev:2000yf,Metsaev:2000yu} and of the
associated phase-space superstring equations of motion.

In section 4 we will   first relate the discussions in sections 2
and 3 by  representing  the light-cone gauge equations  for the
physical string degrees of freedom in  the $\su(2,2|4)$
supermatrix form. This will be done explicitly for a consistent
subsector of solutions with bosonic fields from AdS$_3 \times
$S$^3$ supplemented with 2+2 fermionic fields. Having found the
matrix form of the dynamical equations of motion we will be able
to identify explicitly the corresponding Lax connection and the
associated monodromy matrix, thus demonstrating  how integrability
of the bosonic model generalizes to the presence of fermions. We
shall then find a diagonalization of  the monodromy matrix and the
associated integrals of motion.

In section 5 we shall further specify the discussion of section 4
to an even  smaller subsector of classical configurations AdS$_3
\times $S$^1$  and explicitly relate  the  commuting Cartan
charges associated to nonabelian Noether charge of $\psu(2,2|4)$
to the kinematical generators of symmetries of the light-cone
 gauge superstring. We expect the same relations to hold in the full
 $\AdS$ model. We shall further use our reduced model to investigate
the leading asymptotics of the Lax connection around the branch
cut singularity in spectral parameter. In particular, we will find
that, as in the purely bosonic case \cite{Ka,Arutyunov:2004yx},
the leading asymptotics  turns out to coincide with one of the
global charges which is proportional to the central Dynkin label
of the corresponding $\su(4)$ representation.

In Appendix A  we will give some explicit representations for
various matrices used in the main text. In Appendix B we shall
present the form of the  classical superstring equations reduced
down to the AdS$_3 \times $S$^1$  sector. In Appendix C we shall
review, following \ci{Arutyunov:2004yx}, a method  that allows one
to obtain the leading asymptotics of the Lax connection in the
bosonic $\AdS$ model.

%%%%%%%%%%%%%%%%%%%%%%%%%%%%%%%%%%%%%%%%%%%%%%%%%%%%%%%%%%%%%%%%%%%%%%%%%%
\section{Superstring in $\AdS$ as a supercoset  sigma-model}
%%%%%%%%%%%%%%%%%%%%%%%%%%%%%%%%%%%%%%%%%%%%%
%String theory of $\AdS$ is invariant with respect to the
%superconformal group PSU(2,2$|$4). This fact plays a crucial role
%in constructing .... EOms are rather complicated and to understand
%themn we have to relate with representation theory.
The type IIB Green-Schwarz superstring on the $\AdS$ background
can be defined as a non-linear sigma-model with the following
target space \ci{MT}
 \bea \label{coset} \frac{{\rm PSU}(2,2|4)}{{\rm
SO}(4,1)\times {\rm SO}(5)}\, . \eea The supergroup ${\rm
PSU}(2,2|4)$ with the Lie superalgebra $\psu(2,2|4)$ acts as an
isometry group of the $\AdS$ superspace. We therefore start this
section with recalling the necessary facts about the superalgebra
$\psu(2,2|4)$.

The superalgebra $\su(2,2|4)$ is spanned by $8\times 8$ matrices
$M$ which can be written in terms of $4\times 4$ blocks as \bea
M=\left(
\begin{array}{cc}
  A & X \\
  Y & D
\end{array} \right)\, .
\eea These matrices are required to have vanishing supertrace
${\rm str}M={\rm tr}A-{\rm tr}D=0$ and to satisfy the following
reality condition \bea \label{real} HM+M^{\dagger}H=0\, . \eea For
our  purposes it is convenient to pick up the hermitian
matrix $H$ to be of the form \bea  H=\left(
\begin{array}{rr}
  \Sigma & 0 \\
  0 & -\mathbb{I}
\end{array} \right)\, , ~~~~~~~\Sigma=\left(
\begin{array}{cc}
  0 & \mathbb{I} \\
  \mathbb{I} & 0
\end{array} \right)\, ,
\eea where $\Sigma$ is the $4\times 4$ matrix and $\mathbb{I}$
denotes the identity matrix of the corresponding dimension. The
matrices $A$ and $D$ are even, and $X,Y$ are odd (linear in
fermionic variables). Since the eigenvalues of $\Sigma$ are
$(1,1,-1,-1)$ the condition (\ref{real}) implies that $A$ and $D$
span the subalgebras $\un(2,2)$ and $\un(4)$ respectively, while
$X$ and $Y$ are related as $Y=X^{\dagger}\Sigma$.  The algebra
$\su(2,2|4)$ also contains the $\un(1)$ generator $i\mathbb{I}$ as
it obeys eq.(\ref{real}) and has zero supertrace.

Thus, the bosonic subalgebra of $\su(2,2|4)$ admits the following
decomposition \bea \su (2,2)\oplus \su(4)\oplus \un(1)\, .\eea
Omitting the $\un(1)$ generator one obtains the superalgebra
$\psu(2,2|4)$ we are interested in. It is, however,  important to
note that $\psu(2,2|4)$ {\it can not be realized} as an $8 \times
8$  matrix superalgebra. As we will see this fact becomes
significant if we try to construct the Lax representation for
string equations of motion in the matrix form. The point is that
even if we require the matrices $M$ to be traceless, {\it i.e.}
omit the $\un(1)$ part, it will reappear again through the
commutator of $M's$: \bea [M_1,M_2]=M_3+i\mathbb{I}\Lambda,
~~~~~~~\Lambda\in \mathbb{R}\, . \eea Thus, it makes sense to
define $\psu(2,2|4)$ as the quotient algebra of $\su(2,2|4)$ where
any two elements are considered to be identical if they differ as
matrices only by the identity part.

The superalgebra $\su(2,2|4)$ has a $\mathbb{Z}_4$ grading
$$
M=M^{(0)}\oplus M^{(1)}\oplus M^{(2)}\oplus M^{(3)}
$$
defined by the automorphism $M\to \Omega(M)$ with \bea
\label{Omega} \Omega(M)= \left(
\begin{array}{rr}
  JA^tJ & -JY^tJ \\
  JX^tJ & JD^tJ
\end{array} \right)\, ,
 \eea
where we choose the $4\times 4$ matrix $J$ to be
\begin{equation}
J={\scriptsize\left(
\begin{array}{cccc}
  0 & -1 & 0 & 0 \\
  1 & 0 & 0 & 0 \\
   0 & 0 & 0 & -1 \\
   0 & 0 & 1 & 0
\end{array} \right)\, .}
\end{equation}
The space $M^{(0)}$ is,  in fact,  the ${\rm so(4,1)}\times {\rm
so}(5)$ subalgebra, and the subspaces $M^{(1,3)}$  contain odd
fermionic variables.

Consider now a group element $g$ belonging to ${\rm PSU}(2,2|4)$
and construct the following current \bea \label{la} A=-g^{-1}{\rm
d}g=\underbrace{A^{(0)}+A^{(2)}}_{\rm even
}+\underbrace{A^{(1)}+A^{(3)}}_{\rm odd}\, . \eea Here we
exhibited the $\mathbb{Z}_4$ decomposition of the current. By
construction,  this current has zero-curvature. Let us define \bea
Q=A^{(1)}+A^{(3)} \ , \ \ \ \ \ \ \ \ \ \ Q'=A^{(1)}-A^{(3)}\ ,
\eea and  choose  for $g$ a representative from the coset
(\ref{coset}). Then, as was shown in \cite{Bena}, the $\AdS$
string equations of motion following from the action of \ci{MT}
(which
 includes a Wess-Zumino type term, see also \ci{Roiban:2000yy})
 can be
written in the form \bea \nonumber
&&\pa_{\a}(\gamma^{\a\b}A_{\b}^{(2)})-\gamma^{\a\b}[A_{\a}^{(0)},A_{\b}^{(2)}]
-\frac{1}{2}\epsilon^{\a\b}[Q_{\a},Q'_{\b}]=0\, ,  \\
\label{eom} &&\gamma_{[\a
\rho}\epsilon^{\rho\gamma}Q_{\gamma}A^{(2)}_{\b]} +A^{(2)}_{[\a}
\gamma_{\b]\rho}\epsilon^{\rho\gamma}Q_{\gamma}
-A^{(2)}_{[\a}Q'_{\b]}-Q'_{[\a}A^{(2)}_{\b]}=0 \, ,\\
\nonumber
&&\gamma_{[\a
\rho}\epsilon^{\rho\gamma}Q'_{\gamma}A^{(2)}_{\b]} +A^{(2)}_{[\a}
\gamma_{\b]\rho}\epsilon^{\rho\gamma}Q'_{\gamma}
-A^{(2)}_{[\a}Q_{\b]}-Q_{[\a}A^{(2)}_{\b]}=0 \, .
 \eea
Here we use the convention $\epsilon^{\tau\sigma}=1$, the bracket
$[.,.]$ stands for antisymmetrization of indices and
$\gamma^{\a\b}= h^{\a\b} \sqrt {h}$ is the Weyl-invariant
combination of the metric on  the string
world-sheet.

As in the case of the flat-space   Green-Schwarz  action \ci{GS},
these equations should be supplemented with the Virasoro
constraints which arise upon varying the action w.r.t. the
world-sheet metric. The second two equations include the fermionic
$\kappa$-symmetry constraints (generating the $\kappa$-symmetry
transformations). A decomposition of the full system of
constraints on the first and second class is presently unknown,
and this remains a major obstacle for covariant Hamiltonian
treatment.

\medskip

The integrability  properties of the system (\ref{eom}) were
recently investigated in \cite{Bena,Das}. In
particular, in \cite{Bena} the Lax (zero-curvature)
representation for the system (\ref{eom}) was found. It is based
on the Lax connection $\mathscr{L}$ with components
which have the structure
 \bea
\mathscr{L}_{\a}=\ell_0 A_{\a}^{(0)}+\ell_1 A_{\a}^{(2)}
+\ell_2\gamma_{\a\b}\epsilon^{\beta\rho}A_{\rho}^{(2)} +\ell_3
Q_{\a}+\ell_4 Q'_{\a}\, , \eea where $\ell_i$ are constants.
 The connection $\mathscr{L}$ should
have zero curvature \bea \label{zc}
\pa_{\a}\mathscr{L}_{\beta}-\pa_{\b}\mathscr{L}_{\a}-[\mathscr{L}_{\a},\mathscr{L}_{\b}]=0
\eea as a consequence of the dynamical equations (\ref{eom}) and
of the flatness of $A_{\a}$, and this requirement allows one to
determine $\ell_i$. If we fix $\ell_0$ and $\ell_1$ to be the same
as in the parent bosonic coset model
$$\ell_0=1, ~~~~~~\ell_1=\frac{1+\lambda^2}{1-\lambda^2}\, ,$$
where $\lambda$ is a spectral parameter (not to be confused with
`t Hooft coupling or square of string tension!), then for the
remaining $\ell_i$ we find four different solutions which we group
in two pairs:
\begin{itemize}
\item {\sl First pair}
\bea \label{fp1} \ell_2&=&\frac{2\lambda}{1-\lambda^2}\, ,
~~~~~\ell_3=\frac{1}{\sqrt{1-\lambda^2}}\,
,~~~~~~~\ell_4=\frac{\lambda}{\sqrt{1-\lambda^2}} \, ,
\\
\nonumber \ell_2&=&\frac{2\lambda}{1-\lambda^2}\, ,
~~~~~\ell_3=-\frac{1}{\sqrt{1-\lambda^2}}\,
,~~~~~\ell_4=-\frac{\lambda}{\sqrt{1-\lambda^2}} \, . \eea
\item {\sl Second pair}
\bea \label{sp1} \ell_2&=&-\frac{2\lambda}{1-\lambda^2}\, ,
~~~~~\ell_3=\frac{1}{\sqrt{1-\lambda^2}}\, ,
~~~~~~~\ell_4=-\frac{\lambda}{\sqrt{1-\lambda^2}} \, ,
\\
\nonumber \ell_2&=&-\frac{2\lambda}{1-\lambda^2}\, ,
~~~~~\ell_3=-\frac{1}{\sqrt{1-\lambda^2}}\, ,
~~~~~\ell_4=\frac{\lambda}{\sqrt{1-\lambda^2}}\, . \eea
\end{itemize}
%\bea
%\mathscr{L}_{\a}=A_{\a}^{(0)}+\frac{1+\lambda^2}{1-\lambda^2}A_{\a}^{(2)}
%+\frac{2\lambda}{1-\lambda^2}\gamma_{\a\b}\epsilon^{\beta\rho}A_{\rho}^{(2)}
%+\sqrt{\frac{1}{1-\lambda^2}}Q_{\a}+\sqrt{\frac{\lambda^2}{1-\lambda^2}}Q'_{\a}\,
%, \eea where $\lambda$ is a spectral parameter. One can show then
%that the condition of zero-curvature for $\mathscr{L}$ \bea
%\label{zc}
%\pa_{\a}\mathscr{L}_{\beta}-\pa_{\b}\mathscr{L}_{\a}-[\mathscr{L}_{\a},\mathscr{L}_{\b}]=0
%\eea is equivalent to equations (\ref{eom}).
Two different solutions in each pair reflect the fact that the
dependence on the spectral parameter has two branch cut
singularities at $\lambda=\pm 1$. Going around one of these points
changes the sign in front of $\ell_3$ and $\ell_4$ but it does not
spoil the zero-curvature condition. The two pairs are related by
the identification  $\lambda\to -\lambda$.

When $Q=Q'=0$ the Lax connection reduces to that of the bosonic
model. In the bosonic  case the one-parameter family of the flat
connections allows one to define the monodromy matrix ${\rm
T}(\lambda)$ which is the path-ordered exponential of the Lax
component $\mathscr{L}_{\s}$: \bea {\rm
T}(\lambda)=\mathscr{P}\exp\int_0^{2\pi}{\rm d}\s
\mathscr{L}_{\s}(\lambda)\, . \eea The eigenvalues of ${\rm
T}(\lambda)$ generate infinite set of integrals of motion upon
expansion in $\lambda$ and can be used to study  the spectral
properties of the model. In section 4 we will investigate to which
extent the monodromy matrix can be used in the theory with
fermions.

To derive the classical Bethe equations describing the finite-gap
solutions of the string sigma-model \cite{Ka,bksa} one has to
investigate the asymptotic properties of the Lax connection and
the associated monodromy around the {\it regular} points
$\lambda=0$ and $\lambda=\infty$. These asymptotics must be
related to the global charges of the model;  the latter thus enter
as parameters of the spectral problem.

Suppose we fix a definite branch of $\lambda$ by picking,  for
instance,  the first solution from (\ref{fp1}). We then see that
at $\lambda=0$ the Lax connection reduces to $A_{\a}$. This is
inconvenient for studying the asymptotic behavior of monodromy
around $\lambda=0$. On the other hand,  the Lax equation
(\ref{zc}) is invariant w.r.t. the gauge transformations $$\L\to
\L'=h\L h^{-1}+{\rm d}hh^{-1}\, .$$ This freedom can be used to
gauge away the constant part of $\L$: one has to take $h=g$. Let
us define $a^{({i})}=gA^{(i)}g^{-1}$. The dual current
$\tilde{A}=-{\rm d}gg^{-1}$, which is the inhomogeneous part of
the gauge transformation we perform, can be represented as
 \bea
\tilde{A}=gAg^{-1}=g(A^{(0)}+A^{(1)}+A^{(2)}+A^{(3)})g^{-1}=a^{(0)}+
a^{(1)}+a^{(2)}+a^{(3)}\, . \eea
Then  the result
of the gauge transformation on $\L$ can be written in the form
\bea \label{newLax} \mathscr{L}_{\a}=\ell_0 a_{\a}^{(0)}+\ell_1
a_{\a}^{(2)}
+\ell_2\gamma_{\a\b}\epsilon^{\beta\rho}a_{\rho}^{(2)} +\ell_3
q_{\a}+\ell_4 q'_{\a}\, , \eea where
\bea q=gQg^{-1}\ , \ \ \ \ \ \    q'=gQ'g^{-1}\ , \eea
and $\ell_i$ are now given by \bea \nonumber \ell_0=0\, ,
~~~~~\ell_1=\frac{2\lambda^2}{1-\lambda^2}\, , ~~~~~
\ell_2&=&\frac{2\lambda}{1-\lambda^2}\, ,
~~~~~\ell_3=\frac{1-\sqrt{1-\lambda^2}}{\sqrt{1-\lambda^2}}\,
,~~~~~\ell_4=\frac{\lambda}{\sqrt{1-\lambda^2}} \, . \eea
Expanding this connection around zero  \bea \L_{\a}= \label{zero}
\lambda {\cal L}_{\a} +\ldots \eea we discover that the leading
term ${\cal L}_{\a}$ is \bea {\cal
L}_{\a}=2\gamma_{\a\b}\epsilon^{\beta\rho}a_{\rho}^{(2)}+q'_{\a}
\, . \eea
The zero-curvature condition is satisfied at every
order in $\lambda$;  at order $\lambda$ it gives
 \bea
\pa_{\a}{\cal L}_{\beta}-\pa_{\beta}{\cal L}_{\a}=0 \,
~~~~\Longrightarrow~~~~ \pa_{\a}\Big(\epsilon^{\a\beta}{\cal
L}_{\beta}\Big)=0\,  \eea which is obviously the conservation
equation for a non-abelian current
 \bea \label{cc}
J^{\a}=\epsilon^{\a\b}{\cal
L}_{\beta}=\gamma^{\a\b}a_{\b}^{(2)}+\frac{1}{2}\epsilon^{\a\b}q'_{\a}\,
. \eea This current
 is nothing else but the Noether current of
the global $\psu(2,2|4)$ symmetry of the model. Therefore, the
component $\L_{\s}$ integrated over $\s$  coincides with the
global conserved charge of $\psu(2,2|4)$.

Now suppose we start with  the second solution of (\ref{fp1}). When
$\lambda\to 0$ the Lax connection does not   anymore reduce  to
$A_{\a}$. Still, the constant connection arising in this limit
has zero curvature and,  therefore,  can be gauged away with some
appropriate element $h$. After this  gauge transformation we
find the same type of expansion as in (\ref{zero}) and, as a
consequence, a new non-abelian conserved current.
Our theory has,
however,  the unique non-abelian conserved current corresponding to
the global $\psu(2,2|4)$ symmetry. Therefore,
 the new current should coincide (up to a constant
multiple)  with $KJ^{\a}K^{-1}$,
where $K$ is some constant ({\it i.e.} $\tau$- and
$\sigma$-independent) element.

The analysis of the expansion around $\lambda=\infty$ then goes in a
similar fashion. Expanded around this point the Lax connection has a
constant piece which can be gauged away. After this is done,  at
order $1/\lambda$ one obtains a non-abelian conserved current,
which up to the freedom discussed above,  should be equivalent to
the global $\psu(2,2|4)$ current.

One can also analyze the behavior of the Lax connection and the
monodromy around {\it singular} points $\lambda\to \pm 1$. We
postpone  this  till section 4.

\vskip 0.5cm

Having discussed the generic features of the supercoset  model
let us emphasize
 that both the equations of motion (\ref{eom}) and the
condition of zero curvature (\ref{zc}) hold in the superalgebra
$\psu(2,2|4)$ and, therefore, can not be a priori realized in terms
of matrices. On the other hand, we would like  to have  a matrix
representation for the evolution equations and for the Lax
connection because that would make  the study of the
 spectral properties
of the model fairly easy. As we will see in section 4, working
with matrices will lead to a certain modification of the zero
curvature condition and of the related monodromy matrix.
  This modification is, of course,  entirely due to the
 fermionic degrees of
freedom and does not violate integrability  properties of the model.

Now we are ready to formulate the  basic problem we would like to
address in this paper. The Virasoro constraints do not follow from
the Lax representation (\ref{zc}) and, therefore, provide
additional constraints on our system. However, we  would like  to
know {\it if integrability holds for the physical string}, {\it
i.e.} after we solve all the constraints eliminating all
unphysical degrees of freedom (this also includes fixing the
$\kappa$-symmetry). The lack of the covariant Hamiltonian
formalism makes this clearly an important issue, especially when
it comes to quantization. In section 4 we will verify, by  an
explicit calculation, that the physical string is indeed an
integrable model, at least in the sense that
 it inherits the Lax
representation. To proceed, let us now recall the form of the
superstring equations of motion which arise upon a particular
fixing the $\kappa$- and reparametrization symmetries.

%%%%%%%%%%%%%%%%%%%%%%%%%%%%%%%%%%%%
\section{Superstring in  $\AdS$ in a light-cone gauge}
%%%%%%%%%%%%%%%%%%%%%%%%%%%%%%%%%%%%%%%%%%%%%%%%%%%%%

In this section we  shall review the action and equations of
motion for the $\AdS$ superstring  which arise
 upon fixing the $\kappa$-symmetry on the
world-sheet by the light-cone gauge $\Gamma^+ \theta =0 $
\cite{Metsaev:2000yf,Metsaev:2000yu}.

Let us  parametrize the $\AdS$ metric as
\begin{equation}
ds^2=e^{2 \phi} dx^a dx^a + d\phi^2+ du^M du^M \, ,
\end{equation}
where  the radial coordinate $\phi$ and $x^a$, $a=0,\ldots, 3$,
are  the Poincar\'e coordinates in  $AdS_5$. The five-sphere ${\rm
S}^5$ is parametrized by six embedding coordinates $u^M$,
$M=1,\ldots, 6$, obeying the condition
$$u^Mu^M=1.$$
Let us also  introduce the following combinations of coordinates
%\begin{equation}
$$x^{\pm} = \frac{1}{\sqrt 2} (x^3 \pm x^0)\,\ \ \ \ \ \ \ \ \
x=\frac{1}{\sqrt 2} (x^1 + i x^2)\, , ~~~~~~~
\bar{x}=\frac{1}{\sqrt 2} (x^1 - i x^2)\, .  $$
%\end{equation}
\medskip The Lagrangian describing strings propagating on $\AdS$ in
the $\kappa$-symmetric  gauge is given by \cite{Metsaev:2000yf}
\begin{equation}
{\cal L}={\cal L}_{\rm kin}+{\cal L}_{\rm WZ} \, .
\end{equation}
The kinetic term depends on the world-sheet metric $h_{\a\b}$
(we use $\a=(\tau,\sigma)$ to label the  coordinates on the string
world-sheet)
\begin{eqnarray}
\label{kinlag} {\cal L}_{\rm kin}=-\sqrt{h} h^{\a\b} \big[ e^{2
\phi}(\partial_\a x^+ \partial_\b x^- + \partial_\a x
\partial_\b \bar x)+\frac{1}{2}\partial_\a \phi
\partial_\b \phi + \frac{1}{2} D_\a u^M D_\b u^M \big] \nonumber \\
-\frac{i}{2}\sqrt{h} h^{\a \b} e^{2 \phi} \partial_\a x^+ \big[
\theta^i \partial_\b \theta_i + \theta_i \partial_\b
\theta^i+\eta^i \partial_\b \eta_i + \eta_i \partial_\b \eta^i+ i
e^{2 \phi} \partial_\b x^+ (\eta^2)^2 \big],
\end{eqnarray}
while the Wess-Zumino (topological) term is $h_{\a\b}$-independent
\begin{eqnarray} \la{www}
{\cal L}_{\rm WZ}=\epsilon^{\a \b} e^{2 \phi} \partial_\a x^+
\eta^i \rho_{ij}^M u^M (\partial_\b \theta^j-i \sqrt{2} e^{\phi}
\eta^j
\partial_\b x ) +{\rm h.c.}
\end{eqnarray}
Here
\begin{equation}
D_\a u^M = \partial_\a u^M - 2i \eta_i (R^M)^i_{~j} \eta^j
e^{2\phi} \partial_\a x^+,\hspace{0.3in} R^M=-\frac{1}{2}
\rho^{MN} u^N \, ,
\end{equation}
where the matrices $\rho^{MN}$ are defined in the appendix.
The Lagrangian depends on 16 physical fermionic coordinates,
$\theta^i, \, \eta^i$ and their hermitian conjugates $\theta_i, \,
\eta_i$, where  $i=1,..,4$  is an index of  (anti)fundamental
representation of SU(4).

To obtain the Hamiltonian description  one introduces
the canonical momenta for all the bosonic
variables \bea
\P^{\pm}=\frac{\pa {\cal L}}{\pa \dot{x}^{\mp}}\, ,~~~
\P=\frac{\pa {\cal L}}{\pa \dot{x}}\, , ~~~~ \bar{\P}=\frac{\pa
{\cal L}}{\pa \dot{\bar{x}}}\, ,~~~~ \P_{\phi}=\frac{\pa {\cal
L}}{\pa \dot{\phi}}\, ,~~~
 \P^{M}=\frac{\pa
{\cal L}}{\pa \dot{u}^M}\, . \eea Note that the canonical momenta
$\P^{M}$ satisfy the constraint: \bea \label{con1} \P^Mu^M=0\, .
\eea The bosonic light-cone gauge is imposed
 by requiring the following
two conditions
\begin{equation}
\label{gauge} x^+=\tau,\hspace{0.3in} {\P}^+=p^+\, ,
\end{equation}
where $p^+$ is some non-zero constant.

The Hamiltonian formalism for the light-cone superstring on $\AdS$
was developed in \cite{Metsaev:2000yu} by using the phase
space Lagrangian technique. This approach allows one not only to
find the Hamiltonian for physical fields but also to determine the
world-sheet metric corresponding to the gauge choice
eq.(\ref{gauge}), {\it i.e.} to solve the Virasoro constraints.
Below we shall give a brief summary of the  results of
\cite{Metsaev:2000yu} which are essential for what follows;
for  derivations we refer  to the original work.

Introducing
$\gamma^{\a\b}=\sqrt{h}h^{\a\b}$ with $\det \gamma=-1$,
in the light-cone gauge (\ref{gauge}) we get \cite{Metsaev:2000yu}
\bea \label{metric} \gamma^{\tau\tau}=-p^+e^{-2\phi}\, ,
~~~~~~\gamma^{\s\s}=\frac{1}{p^+}e^{2\phi}\, ,
~~~~~~\gamma^{\tau\sigma}=\gamma^{\sigma\tau}=0\, . \eea The
Hamiltonian density ${\cal H}\equiv -{\P^-}$ is given by
\begin{eqnarray}
\label{H} {\cal H}&=&\frac{1}{2p^+} \left[
2{\P}\bar{\P}+2e^{4\phi} \acute{x}\acute{\bar{x}}
+e^{2\phi}(\P_{\phi}^2+\acute{\phi}^2 + {l^i_j}^2+\acute{u}^M
\acute{u}^M + p^{+ 2}(\eta^2)^2+4p^+ \eta_i
l^i_j \eta^j ) \right] \nonumber\\
&&~+e^{2\phi} \eta^i y_{ij}(\acute{\theta}^j-i \sqrt{2} e^\phi
\eta^j \acute x)+e^{2\phi} \eta_i y^{ij}(\acute{\theta}_j+i
\sqrt{2} e^\phi \eta_j \acute{\bar x}) \, ,
\end{eqnarray}
where we defined
\begin{equation}
\nonumber y_{ij} \equiv \rho_{ij}^M u^M,\hspace{0.3in}y^{ij}
\equiv (\rho^{ij})^M u^M,\hspace{0.3in} l^i_j \equiv \frac{i}{2}
(\rho^{MN})^i_j u^M {\P}^N\, .
\end{equation}
Note that taking into account the constraint (\ref{con1}) we get
$l^i_{\,k}l^k_{\, j}=\frac{1}{4}{\P}^M {\P}^M \delta^i_{\, j}$.

As usual in  the light-cone gauge the field $x^-$ appears to
be unphysical. Its $\sigma$ derivative $\acute{x}^-$ is expressed in terms
of physical fields as
\begin{equation}
\label{sd} \acute{x}^- =-\frac{1}{p^+}\big[
{\P}\acute{\bar{x}}+\bar{\P}\acute{x}+\P_{\phi} \acute{\phi}
+{\P}_M \acute{u}^M+\frac{i}{2}p^+ ( \theta^i \acute{\theta}_i +
\theta_i \acute{\theta}^i+\eta^i \acute \eta_i + \eta_i \acute
\eta^i )\big]\, ,
\end{equation}
while the evolution equation is \begin{eqnarray} \nonumber
\dot{x}^- =&-&\frac{1}{2(p^+)^2} \big[ 2{\P}\bar{\P}+2e^{4\phi}
\acute{x}\acute{\bar{x}}\\
\nonumber &+&e^{2\phi}(\P_{\phi}^2+\acute{\phi}^2 +
\P_M^2+\acute{u}^M \acute{u}^M - p^{+ 2}(\eta^2)^2+4p^+ \eta_i
l^i_j \eta^j ) \big] \\
\label{td} &-&\frac{i}{p^+}e^{2\phi}\eta_i(\rho^{MN})^i{}_j\eta^j
\P^Mu^N -\frac{i}{2}p^+ ( \theta^i \dot{\theta}_i + \theta_i
\dot{\theta}^i+\eta^i \dot{\eta}_i + \eta_i \dot{\eta}^i )\, .
\end{eqnarray}
Since we consider closed strings the zero mode ${\cal V}$ of
$x^{-}$,
$$
{\cal V}=\int_0^{2\pi}\frac{{\rm d}\sigma}{2\pi} \big[
{\P}\acute{\bar{x}}+\bar{\P}\acute{x}+\P_{\phi} \acute{\phi}
+{\P}_M \acute{u}^M+\frac{i}{2}p^+ ( \theta^i \acute{\theta}_i +
\theta_i \acute{\theta}^i+\eta^i \acute \eta_i + \eta_i \acute
\eta^i )\big]\, ,
$$
leads to the residual constraint ${\cal V}=0$ which we leave
unsolved.

Supplying the Hamiltonian with the proper Poisson-Dirac brackets,
the Hamiltonian equations of motion for the physical fields are found
to be:

\vskip 0.5cm {\small \noindent {\sl AdS bosonic fields}
\begin{eqnarray}
%\label{eqmot1}
\nonumber \dot{x} &=& \frac{1}{p^+}\P\,,\ \ \qquad \dot{\bar{x}} =
\frac{1}{p^+}\bar{\P}\ ,  \ \ \ \ \ \ \ \ \dot{\phi} =
\frac{e^{2\phi}}{p^+}\P_{\phi} \ ,
\\
\nonumber \dot{\P} &=& \frac{1}{p^+}\partial_\sigma(e^{4\phi}\x')
-{\rm i}\sqrt{2}\partial_\sigma(e^{3\phi}\eta_i y^{ij}\eta_j)
\ , \\
\nonumber \dot{\bar{\P}} &=&
\frac{1}{p^+}\partial_\sigma(e^{4\phi}\xb') +{\rm
i}\sqrt{2}\partial_\sigma(e^{3\phi}\eta^iy_{ij}\eta^j) \ ,
\\
\nonumber \dot{\P}_{\phi} &=& \frac{1}{p^+}\partial_\sigma
(e^{2\phi}\ph') -\frac{4}{p^+}e^{4\phi}\x'\xb'
\\
\nonumber &-&\frac{e^{2\phi}}{p^+}\Bigl(\P_{\phi}^2+
\ph'^2+\P^M\P^M + \u'^M\u'^M +p^{+2}(\eta^2)^2 + 4p^+\eta_i
l^i{}_j \eta^j \Bigr)
\nonumber\\
\label{abf} &+& e^{2\phi}\eta^i y_{ij} (2\th'^j - 3{\rm
i}\sqrt{2}e^\phi \eta^j\x') +e^{2\phi}\eta_i y^{ij} (2\th'_j +
3{\rm i}\sqrt{2}e^\phi \eta_j\xb')\, .
\end{eqnarray}
\noindent {\sl Sphere bosonic fields}
\begin{eqnarray}
\nonumber \dot{u}^M &=&\frac{e^{2\phi}}{p^+}\P^M -{\rm
i}e^{2\phi}\eta_i(\rho^{MN})^i{}_j\eta^j u^N \ ,
\\
\nonumber
 \dot{\P}^M & =& -\frac{e^{2\phi}}{p^+}u^M \P^N\P^N
+\frac{1}{p^+}v^{MN}\partial_\sigma(e^{2\phi}\u'^N) -{\rm
i}e^{2\phi}\eta_i(\rho^{MN})^i{}_j\eta^j \P^N
\\
\label{sbf} &+& e^{2\phi}v^{MN}\eta^i\rho_{ij}^N (\th'^j-{\rm
i}\sqrt{2}e^\phi \eta^j \x') +e^{2\phi}v^{MN}\eta_i(\rho^N)^{ij}
(\th'_j+{\rm i}\sqrt{2}e^\phi \eta_j \xb')\, .
\end{eqnarray}
\noindent{\sl Fermions}
\begin{eqnarray}
\nonumber \dot{\theta^i} &=& -\frac{\rm
i}{p^+}\partial_\sigma(e^{2\phi}y^{ij}\eta_j) \ , \ \ \ \ \ \
\dot{\theta_i} = -\frac{\rm
i}{p^+}\partial_\sigma(e^{2\phi}y_{ij}\eta^j) \ ,
\\
\nonumber \dot{\eta}^i &=& e^{2\phi} \Bigl[{\rm i}\eta^2\eta^i
-\frac{2\rm i}{p^+}(l\eta)^i +\frac{\rm i}{p^+} y^{ij}(\th'_j +
{\rm i}2\sqrt{2}e^\phi\eta_j\xb')\Bigr]\ ,
\\
\label{fermion} \dot{\eta}_i &=& e^{2\phi} \Bigl[-{\rm
i}\eta^2\eta_i +\frac{2\rm i}{p^+}(\eta l)_i +\frac{\rm
i}{p^+}y_{ij}(\th'^j-{\rm i}2\sqrt{2}e^\phi\eta^j\x')\Bigr]\ .
\end{eqnarray}
} \medskip

\noindent Here the equations of motion for fields parametrizing
the five-sphere involve the following tensor
\begin{equation}
v^{MN} \equiv \delta^{MN} -u^M u^N\, .
\end{equation}
and we use the notation $\eta^2\equiv \eta^i\eta_i$. We also do
not distinguish between the upper and lower indices $M,N$, i.e.
use the convention $\P_M \equiv \P^M$. The fermionic variables
obey the following hermitian conjugation rule:
$\eta_i^\dagger=\eta^i$, $\theta_i^\dagger=\theta^i$ and
$(f_1f_2)^\dagger=f_2^\dagger f_1^\dagger$
 if $f_1,f_2$ are
fermions. This implies that $\eta^2$ and $\theta^2$ are hermitian even
variables.

Note that we  did  not
attempt to replace  the time derivatives of fermions by the
corresponding canonical momenta which are determined  by solving
the following second class constraints (and similar
 ones  for $\eta^i$,
$\eta_i$)
\be
 \P_{\theta^i}+\frac{i}{2}p^+\theta_i=0\,,
\qquad
 \P_{\theta_i}+\frac{i}{2}p^+\theta^i=0 \ ,
\ee where $\P_{\theta^i}$, $\P_{\theta_i}$ are the canonical
momenta for the fermionic variables.
This will  not be needed
for our present  purposes.

%%%%%%%%%%%%%%%%%%%%%%%
\section{Integrability on   physical subspace}
%%%%%%%%%%%%%%%%%%%%%%%%%%%%%%%%%%%%%%5

Let us now relate the discussions in sections 2 and 3
by  putting  the light-cone gauge equations  for the
physical degrees
of freedom \rf{abf},\rf{sbf},\rf{fermion} in a matrix form
as in \rf{eom}.
For this  we need to
choose an  appropriate embedding of the  coset
 representative eq.(\ref{coset}) into
the matrix supergroup SU(2,2$|$4).
We shall make the same choice as in
\cite{Metsaev:2000yf} which was used to fix the
light-cone
 $\kappa$-symmetry gauge (and led to the Lagrangian
 \rf{kinlag},\rf{www}  which is quartic in fermions).
We define the $\kappa$-gauge fixed coset representative $g$ as a
product of four elements
\bea \la{ggg}
g=g(x,\theta)g(\eta)g(y)g(\phi)\, ,
\eea where
\begin{eqnarray}
\label{cr}
\begin{array}{ll}
g(x,\theta)=\exp{(x^i {\rm P}^i +
Q)}\, , ~~~~~~~~ & g(\eta)=\exp{(S)},\\
 & \\
 g(y)=\exp{\frac{i}{2}(y^{\mu}\Gamma^{\mu})}\, , & g(\phi)=\exp{(\phi {\rm
 D})}\, .
\end{array}
\end{eqnarray}
Here ${\rm P}^i$, $i=1,\ldots, 4$, and ${\rm D}$ are the
generators of translations and scale transformations respectively.
Together with the Lorentz boosts and special conformal
transformations they form the conformal subalgebra $\su(2,2)$. In
appendix A we give an explicit realization of these generators in
terms of $4\times 4$ matrices and then trivially embed them in
$8\times 8$ matrices to represent the corresponding generators of
$\su(2,2|4)$. The SO(5) Dirac matrices $\Gamma^{\mu}$,
$\mu=1,\ldots, 5$, are also collected in appendix A.

The supercharges $Q$ and $S$ represent the conformal and special
supersymmetries, each of them is expressed in terms  of
 16 independent fermionic
variables $\theta$ and  $\eta$ respectively. We realize them as
the following matrices
\medskip

\begin{equation}
Q=2^{\frac{1}{4}}{\footnotesize \left(
\begin{array}{cccccccc}
  0 & 0 & 0 & 0 & 0 & 0 & 0 & 0 \\
  0 & 0 & 0 & 0 & 0 & 0 & 0 & 0 \\
  0 & 0 & 0 & 0 & \theta^1 & \theta^2 & \theta^3 & \theta^4 \\
  0 & 0 & 0 & 0 &  \theta^5 & \theta^6 & \theta^7 & \theta^8 \\
  \theta_1 & \theta_5 & 0 & 0 & 0 & 0 & 0 & 0 \\
  \theta_2 & \theta_6 & 0 & 0 & 0 & 0 & 0 & 0 \\
  \theta_3 & \theta_7 & 0 & 0 & 0 & 0 & 0 & 0 \\
  \theta_4 & \theta_8 & 0 & 0 & 0 & 0 & 0 & 0
\end{array}\right)
}, \hspace{0.3in}S=2^{\frac{1}{4}}e^{\phi}{\footnotesize \left(
\begin{array}{cccccccc}
  0 & 0 & 0 & 0 & \eta^5 & \eta^6 & \eta^7 & \eta^8 \\
  0 & 0 & 0 & 0 & \eta^1 & \eta^2 & \eta^3 & \eta^4 \\
  0 & 0 & 0 & 0 & 0 & 0 & 0 & 0 \\
  0 & 0 & 0 & 0 & 0 & 0 & 0 & 0 \\
  0 & 0 & \eta_5 & \eta_1 & 0 & 0 & 0 & 0 \\
  0 & 0 & \eta_6 & \eta_2 & 0 & 0 & 0 & 0 \\
  0 & 0 & \eta_7 & \eta_3 & 0 & 0 & 0 & 0 \\
  0 & 0 & \eta_8 & \eta_4 & 0 & 0 & 0 & 0
\end{array}\right)
}
\end{equation}
\medskip

\noindent The scaling factors in front of the matrices are
introduced for further convenience. Fixing the $\kappa$-symmetry
as in \cite{Metsaev:2000yf} amounts to putting to zero the
following fermionic variables \bea
\theta^5=...=\theta^8=\eta^5=...=\eta^8=0 \eea and also their
conjugate partners with lower indices.

The physical variables $x^i,\phi$, $y^{\mu}$ and
$\theta_i,\theta^i$, $\eta_i,\eta^i$ parametrize the
$\kappa$-gauge fixed coset representative (\ref{coset}). The
coordinates $x^i$ are given in terms of light-cone variables by
\bea \nonumber x^0&=&\frac{1}{\sqrt{2}}(x^--x^+)\, ,~~~~~~
x^1=\frac{1}{\sqrt{2}}(x^--x^+)\, , \\
\nonumber x^2&=&-\frac{1}{\sqrt{2}}(x+\bar{x})\ ,~~~~~~~
x^3=-\frac{i}{\sqrt{2}}(x-\bar{x})\, . \eea The coordinates
$y^{\mu}$ parametrize the five-sphere. In what follows it is
convenient to use the six embedding coordinates $u^M$ (because
they enter the equations of motion (\ref{sbf})) which are
expressed through $y^{\mu}$ as \bea u^6=\cos y\, ,
~~~~~~~u^{\mu}=\frac{y^{\mu}}{y}\sin y\, , \ \ \ \ \ \
y=\sqrt{(y^1)^2+\ldots (y^5)^2}\ .  \eea Now we can use the
$8\times 8$ matrix $g$ in \rf{ggg} to construct the current
eq.(\ref{la}) and find the corresponding $\mathbb{Z}_4$
decomposition with respect to $\Omega$ (\ref{Omega}). For even
elements we have \bea \nonumber
A^{(0)}&=&\frac{1}{4}\big(A+\Omega(A)+\Omega^2(A)+\Omega^3(A)
\big)\, ,\\
\la{ee}
A^{(2)}&=&\frac{1}{4}\big(A+i^2\Omega(A)+i^4\Omega^2(A)+i^6\Omega^3(A)
\big) \eea and for odd \bea \nonumber
A^{(1)}&=&\frac{1}{4}\big(A+i^3\Omega(A)+i^6\Omega^2(A)+i^9\Omega^3(A)
\big) \, , \\
\la{oo}
A^{(3)}&=&\frac{1}{4}\big(A+i\Omega(A)+i^2\Omega^2(A)+i^3\Omega^3(A)
\big)\, .
 \eea
Using the expression
for  the world-sheet metric (\ref{metric}),
 the l.h.s.
of the equations of motion (\ref{eom})
thus  acquires the
 following  form in the light-cone gauge
\bea \label{E1} \nonumber &&{\rm E}_1=\frac{1}{p^+}
\pa_{\s}(e^{2\phi}A_{\s}^{(2)})-p^+\pa_{\tau}(e^{-2\phi}A_{\tau}^{(2)})
+\frac{1}{p^+}e^{2\phi}[A_{\s}^{(2)},A_{\s}^{(0)}]
\\
&&~~~~~~~~~~~-p^+e^{-2\phi}[A_{\tau}^{(2)},A_{\tau}^{(0)}]-\frac{1}{2}[Q_{\tau},Q'_{\s}]-\frac{1}{2}[Q'_{\tau},Q_{\s}]\,
,
\\
\label{E2} && {\rm E}_2=\frac{1}{p^+}e^{2\phi}
[A_{\s}^{(2)},Q_{\s}]-p^+e^{-2\phi}[A_{\tau}^{(2)},Q_{\tau}]+[A_{\s}^{(2)},Q'_{\tau}]
-[A_{\tau}^{(2)},Q'_{\s}]\, , \\
\label{E3} && {\rm E}_3=\frac{1}{p^+}e^{2\phi}
[A_{\s}^{(2)},Q'_{\s}]-p^+e^{-2\phi}[A_{\tau}^{(2)},Q'_{\tau}]+[A_{\s}^{(2)},Q_{\tau}]
-[A_{\tau}^{(2)},Q_{\s}]\, . \eea Next,  let us compute the
current  $A$  \rf{la} constructed from the coset representative
(\ref{ggg}),\rf{cr}, find  the corresponding projections $A^{(0)},
\ldots, A^{(3)}$  \rf{ee},\rf{oo} and  plug them into
eqs.(\ref{E1})-(\ref{E3}).  We  can then use the light-cone gauge
equations of motion (\ref{sd})-(\ref{fermion}) for the
$\kappa$-fixed Hamiltonian
 and the bosonic light-cone  gauge condition
(\ref{gauge}) to express the result in terms of
the  physical fields
only.
%(\ref{td}), (\ref{abf}),(\ref{sbf}),
\medskip

Dealing with the full  $\AdS$ model appears
to be rather complicated,
  so we shall restrict our consideration
%we were able to make a progress for the
to a consistent  subsector of solutions of equations of motion
which we shall
call ${\rm AdS}_3\times {\rm S}^3$.
By a consistent reduction to a subsector we  mean
that if we put some fields to zero then they will remain zero
as a consequence of  their Hamiltonian equations.
One can show that it is
 a consistent reduction of the string equations
(\ref{abf}),(\ref{sbf}) and (\ref{fermion}) to
switch off the following fields
\begin{eqnarray}
x=\bar{x}=\P=\bar{\P}=u^5=u^6=\P^5=\P^6=0\, , \\
\eta_1=\eta^1=\eta_2=\eta^2=\theta_3=\theta^3=\theta_4=\theta^4=0\,
\la {p}.
\end{eqnarray}
We are then left with four coordinates $u^M$ parametrizing a
three-sphere and the radial  field $\phi$ which together with
$x^{\pm}$ (which are eliminated by our gauge choice)
 describe the ${\rm AdS}_3$ space. We note that
a further reduction to ${\rm AdS}_3\times {\rm S}^1$ is
possible by setting
\bea
u^1=u^4=\P^{1}=\P^{4}=0\, , \\
\eta_3=\eta^3=\theta_2=\theta^2\, . \la{g}
 \eea
It is worth emphasizing  that calling the reduced models as  ${\rm
AdS}_3\times {\rm S}^3$ or ${\rm AdS}_3\times {\rm S}^1$ we refer
to {\it dimensional reduction of the bosonic string}. The
corresponding reduction
 of the fermionic variables is  then
dictated by the equations of motion. \footnote{Note that the
light-cone supersymmetry generators can also be consistently
truncated.} The remaining fermions, a priori, need not be the same
 fermionic variables
which we would get if we would  start directly
 with the superstring in
six \ci{mtsix}
 or four dimensions:
 in our procedure we first
impose the gauge and then perform the reduction,
 which is apparently not the
same as to use the $\kappa$-symmetric
Green-Schwarz superstring in
lower dimension and fix the gauge there.

\medskip

Restricting to the
 ${\rm AdS}_3\times {\rm S}^3$ sector,
  expressing
eqs.(\ref{E1}),(\ref{E2}) in
terms of the light-cone fields and
using their equations (\ref{abf}),(\ref{sbf}) and (\ref{fermion})
we find that
\bea {\rm E}_2={\rm
E}_3=0 \, ,\eea
%while for ${\rm E}_1$ we obtain
\bea \label{anom}
{\rm E}_1=i\Lambda\mathbb{I}_{8\times 8}\, , \eea
 where \bea
\Lambda=p^+\pa_{\tau}(\eta_3\eta^3+\eta_4\eta^4)\equiv
p^+\pa_{\tau}(\eta_i\eta^i)\, , ~~~~~~i=3,4\, .\la{aan}
 \eea
Since ${\rm E}_1$ is non-vanishing only modulo  a unit matrix, we
conclude that the dynamical string equations
 for connections in   $\psu(2,2|4)$ are
exactly satisfied on solutions of
 the $\kappa$-symmetry and
Virasoro constraints. In other words, we have obtained a
representation of the light-cone equations of motion
(\ref{abf}),(\ref{sbf}) and (\ref{fermion}) in terms of the
dynamical equations (\ref{E1}),(\ref{E2}) and (\ref{E3}) imposed
on $8\times 8$ matrices from $\su(2,2|4)$.

%Realizing the dynamical equations in terms of
%matrices from $\su(2,2,|4)$ we get,  however,
%an  additional term in
%eq.(\ref{anom}) proportional to the identity matrix.
%According to
%the logic of our construction, this term is absent in the
%superalgebra $\psu(2,2|4)$.

As we will see later the ``anomalous'' $\Lambda $ term in
\rf{anom} (present in  matrix $\su(2,2|4)$ realization but
factored out in the physical
 $\psu(2,2|4)$  case)
 will
not cause any difficulty in studying  the
integrability  properties of
the model by means of a concrete matrix representation.

\medskip

Let us now look at the Lax connection
 (\ref{zc}) with coefficients
(\ref{fp1}), or,  explicitly,
  \bea \nonumber
\mathscr{L}_{\tau}&=&A_{\tau}^{(0)}+\frac{1+\lambda^2}{1-\lambda^2}A_{\tau}^{(2)}
-\frac{2\lambda}{1-\lambda^2}\frac{e^{2\phi}}{p^+}A_{\s}^{(2)}
+\frac{1}{\sqrt{1-\lambda^2}}Q_{\tau}+\frac{\lambda}{\sqrt{1-\lambda^2}}Q'_{\tau}\, ,\\
\nonumber
\mathscr{L}_{\s}&=&A_{\s}^{(0)}+\frac{1+\lambda^2}{1-\lambda^2}A_{\s}^{(2)}
-\frac{2\lambda}{1-\lambda^2}p^+e^{-2\phi}A_{\tau}^{(2)}
+\frac{1}{\sqrt{1-\lambda^2}}Q_{\s}+\frac{\lambda}{\sqrt{1-\lambda^2}}Q'_{\s}\,
. \eea The Lax equation (\ref{zc}) follows from the   two
conditions: the current $A$ is flat and it satisfies equations of
the motion (\ref{eom}). It remains flat when we realize it as a
$\su(2,2|4)$ matrix, but the equations of motion get modified due
to the $\Lambda$-term in \rf{anom}. We should then expect that the
curvature of $\mathscr{L}$ viewed as a matrix in $\su(2,2|4)$ is
no  longer zero. Indeed, by the explicit calculation we find \bea
\label{Lax}
\pa_{\tau}\mathscr{L}_{\sigma}-\pa_{\sigma}\mathscr{L}_{\tau}-[\mathscr{L}_{\tau},\mathscr{L}_{\sigma}]=
i\frac{2\lambda}{1-\lambda^2}\Lambda~ \mathbb{I}\, . \eea Still,
the curvature vanishes when restricted to
 $\psu(2,2|4)$. We conclude,
therefore,
that the {\it physical string
 is an integrable model}.

\bigskip

 Let us now
  proceed with our explicit matrix Lax
representation and define the conserved quantities corresponding
to eq.(\ref{Lax}). Let us compute the time derivative of the
monodromy matrix: \bea \pa_{\tau}{\rm T}=
 \int_0^{2\pi}{\rm d}\s'~
\left(\mathscr{P}\exp\int_{\s'}^{2\pi} \L_{\s}\right)
\pa_{\tau}\L_{\s}(\s',\tau)\left(\mathscr{P}\exp\int^{\s'}_{0}
\L_{\s}\right)\, .\eea We can now use eq.(\ref{Lax}) to rewrite
this as \bea \pa_{\tau}{\rm T}=[\L_{\tau}(0,\tau),{\rm
T}]+i\frac{2\lambda}{1-\lambda^2}{\rm T}\int_0^{2\pi}{\rm d}\sigma
\Lambda(\sigma)\, . \eea

\medskip

\noindent For a generic value of the spectral parameter the
monodromy matrix is diagonalizable by means of a regular group
element $\gg$ and we can write
%(not to be confused of course with $g$ in  \rf{ggg}):
\begin{equation}
{\rm T}=\gg D \gg^{-1}\, ,
\end{equation}
were $D$ is a diagonal $\su(2,2|4)$ matrix.
%%%%%%%%%%%%%%%%%%%%%%%%%%%%%%%%%%%%%%%
%%%%%%%%%%%%%%%%%%%%%%%%%%%%%%%%%%%%%%%%%%%%%%%%%%%%%
Then
 \bea\la{kkk}
\pa_{\tau}D=[\gg^{-1}\mathscr{L}_{\tau}\gg-\gg^{-1}\pa_{\tau}\gg,D]
+i\frac{2\lambda}{1-\lambda^2}D\int_0^{2\pi}{\rm d}\sigma
\Lambda(\sigma)\,  .\eea Using the explicit form of $\Lambda$ in
\rf{aan} this relation can be written as \bea \label{fff}
\pa_{\tau} \JI(\lambda) % \left(\log{D}-i\frac{\lambda}{1-\lambda^2}\mathbb{I}\int_0^{2\pi}{\rm
%d}\sigma\eta_i\eta^i\right)
=[\gg^{-1}\mathscr{L}_{\tau}\gg-\gg^{-1}\pa_{\tau}\gg,\JI(\l)]\, ,
\la{jj}\eea \bea \JI(\lambda)\equiv \exp\left(
-i\frac{2\lambda}{1-\lambda^2}\int_0^{2\pi}p^+ \eta_i\eta^i {\rm
d} \sigma\, \right) D(\l).  \la{iii} \eea In the purely bosonic
case the r.h.s. of eq.\rf{fff} would need to vanish because
$\JI(\l)$ is diagonal while the commutator is off-diagonal and
then  we would obtain an infinite set of conservation laws
generated by $\JI(\lambda)$ upon expansion in the spectral
parameter. In the presence of fermions a matrix $\gg$ which
diagonalizes monodromy can be chosen from ${\rm SU}(2,2|4)$ so
that $\JI(\lambda)$ is an even element. Hence, the commutator  in
\rf{jj} should also vanishes as in the bosonic case. As a
consequence, the quantity $\JI(\lambda)$ in \rf{iii} is
conserved.\footnote{We thank Sergey Frolov for an important
discussion of this point.}

\medskip

At $\lambda=\pm 1$ the Lax connection becomes however singular
implying the essential singularity of the corresponding monodromy
matrix at these points. This case requires special treatment. As
is known \cite{FTa} for the purely bosonic model the asymptotic
expansion of the monodromy around $\lambda=\pm 1$ produce  {\it
local} integrals of motion. Let us now show that in the present
fermionic case the standard asymptotic analysis of the Lax
connection around $\lambda=\pm 1$ does not apparently give the
local conservation laws.

%Let now perform the standard asymptotic analysis In the fermionic
%case the points $\lambda=\pm 1$ are the branch cut singularities
%of the Lax connection. Let us now perform a local analysis of the
%Lax equation around, {\it e.g.}, $\lambda=1$. As we will see, the
%local conservation laws present in the bosonic theory will be
%apparently lost.
\medskip

Expanding the  Lax connection around $\lambda =1$
we get  (we assume
$0<\lambda<1$) \bea
\L_{\a}=\frac{1}{1-\lambda}\L_{\a}^{(0)}+\frac{1}{\sqrt{1-\lambda}}\L_{\a}^{(1)}+
\L^{(2)}_{\a}+\ldots \, , \eea
where the matrices $\L^{(0)}$ and $\L^{(2)}$ are even
while $\L^{(1)}$
is odd. Substituting this into eq.(\ref{Lax}) we see that vanishing of
the first three most singular terms requires the fulfillment of the
following equations: \bea \label{1}
&&\frac{1}{(1-\lambda)^2}:~~~~~~[\L_{\a}^{(0)},\L_{\b}^{(0)}]=0\\
\label{2}
&&\frac{1}{(1-\lambda)^{\frac{3}{2}}}:~~~~~~[\L_{\a}^{(0)},\L_{\b}^{(1)}]-[\L_{\b}^{(0)},\L_{\a}^{(1)}]=0\\
\nonumber
&&\frac{1}{(1-\lambda)}:~~~~~~~\pa_{\a}\mathscr{L}_{\b}^{(0)}
-\pa_{\b}\mathscr{L}_{\a}^{(0)}-[\L_{\a}^{(1)},\L_{\b}^{(1)}]-\\
\label{3} &&~~~~~~~~~~~~~~~~~~~~~~~~~~~~~
-[\L_{\a}^{(0)},\L_{\b}^{(2)}]-[\L_{\a}^{(2)},\L_{\b}^{(0)}]
=i\Lambda \mathbb{I}\, . \eea The first condition tells us that
$\L_{\tau}^{(0)}$ and $\L_{\s}^{(0)}$ commute with each
other\footnote{This follows from the explicit expression for $\L$
which implies that $\L_{\s}^{(0)}=-p^+e^{-2\phi}\L_{\tau}^{(0)}$.}
and,  therefore,
 can be simultaneously diagonalized by a similarity
transformation \bea \L_{\tau}^{(0)}=\gg D_{\tau}\gg^{-1}\, ,\ \ \ \ \ \
~~~~\L_{\s}^{(0)}=\gg D_{\s}\gg^{-1}\, \eea with some {\it even}
element $\gg$. The second equation (\ref{2}) then becomes \bea
\nonumber
[D_{\s},\gg^{-1}\L_{\tau}^{(1)}\gg]-[D_{\tau},\gg^{-1}\L_{\s}^{(1)}\gg]=0\,
,
 \eea
while eq.(\ref{3}) reduces to \bea  \nonumber &&
\pa_{\tau}D_{\s}-\pa_{\s}D_{\tau}+[\gg^{-1}\pa_{\tau}\gg
-\gg^{-1}\L_{\tau}^{(2)}\gg,D_{\s}] \\
&&~~~~~~~~~~~~ -[\gg^{-1}\pa_{\s}\gg
-\gg^{-1}\L_{\sigma}^{(2)}\gg,D_{\tau}]-\gg^{-1}[\L_{\a}^{(1)},
\L_{\b}^{(1)}]\gg =i\Lambda\mathbb{I}\, . \eea The commutators of
the even elements involving the diagonal matrices $D_{\tau}$ and
$D_{\sigma}$ do not have diagonal part. Therefore, projecting the
last equation on the diagonal part we obtain \bea \label{basic}
\pa_{\tau}D_{\s}-\pa_{\s}D_{\tau}=
\Big(\gg^{-1}[\L_{\tau}^{(1)},\L_{\sigma}^{(1)}]\gg\Big)_{\rm
diag} +i\Lambda\mathbb{I}\, . \eea We see that \bea \label{wbc}
I=\int_0^{2\pi}\frac{{\rm d\s}}{2\pi}D_{\s} \eea is not conserved
as it would be for the  bosonic model. The $\Lambda$-term does not
cause any problem as it appears to be a time-derivative.
Non-conservation of the current is due to the fermionic source
$\Big(\gg^{-1}[\L_{\tau}^{(1)},\L_{\sigma}^{(1)}]\gg\Big)_{\rm
diag}$ which a priori cannot be written in the form \bea
\label{factor}
\Big(\gg^{-1}[\L_{\tau}^{(1)},\L_{\sigma}^{(1)}]\gg\Big)_{\rm
diag}=\pa_{\tau}V_{\s}-\pa_{\s}V_{\tau} \eea for some  $V_{\s}$
and $V_{\tau}$ which are {\it local} functions of $\tau$ and
$\sigma$. Note that representing the fermionic source in the form
(\ref{factor}) should not involve equations of motion, since the
equations of motion make  eq.(\ref{basic}) into  an identity.

\medskip

For the case of the  ${\rm AdS}_3\times {\rm S}^3$
sector we found that the
matrix $D_{\s}$ takes the  following
form
 \bea D_{\s}=\frac{i}{2}~{\rm
diag}\Big(-\kappa,-\kappa,+\kappa,+\kappa,-\kappa,-\kappa,+\kappa,+\kappa\Big)+\frac{i}{2}p^+\eta_i\eta^i\mathbb{I}_{8\times
8 }\, , \eea where \bea \label{kappa}
\kappa^2=(\P^M-\acute{u}^M)^2\, . \eea The explicit form of the
matrix $\gg$ which diagonalizes
 $\L_{\a}^{(0)}$ is
given in Appendix A. Remarkably, this matrix does not depend on
fermionic variables. As we have chosen to diagonalize
$\L_{\a}^{(0)}$ with $\gg$ from ${\rm SU}(2,2|4)$ the matrix
$D_{\s}$ has zero supertrace.

\medskip

Let us now relate our general discussion of the monodromy and
associated conservation laws with the present local analysis
around singularity at $\lambda=1$. To this end consider the matrix
$\L_{\sigma}$ and try to diagonalize it with a {\it regular} gauge
transformation: \bea \nonumber {\rm g}={\rm
g_0}+\sqrt{1-\lambda}~{\rm g_1}+\ldots ,~~~~~~~ {\rm g^{-1}}={\rm
g_0^{-1}}-\sqrt{1-\lambda}~{\rm g_0^{-1}g_1g_0^{-1}}+\ldots \eea
This produces an expansion \bea \nonumber &&{\rm
g^{-1}}\L_{\sigma}{\rm g}-{\rm g}^{-1}\pa_{\s}{\rm g}=
\frac{1}{1-\lambda}{\rm g_0^{-1}}\L_{\s}^{(0)}{\rm g_0}+\\
\nonumber &&~~~~~~~~~~~~~ +\frac{1}{\sqrt{1-\lambda}}\Big({\rm
g_0^{-1}}\L_{\s}^{(1)}{\rm g_0}-[{\rm g_0^{-1}g_1},{\rm
g_0^{-1}}\L_{\s}^{(0)}{\rm g_0}] \Big)+\ldots \, .\eea Since we
have chosen an even matrix ${\rm g_0}$ to diagonalize
$\L_{\s}^{(0)}$ the coefficient of the branch cut singularity can
be written in the form \bea \label{sing} {\rm
g_0^{-1}}\L_{\s}^{(1)}{\rm g_0}+[D_{\s}, {\rm g_0^{-1}g_1}]\, .
\eea Obviously, if an even diagonal matrix $D_{\s}$ is
non-degenerate, {\it i.e.} does not have any coinciding elements,
then one can always find some odd supermatrix ${\rm g_1}$ such
that all non-diagonal elements of (\ref{sing}) vanish. In this
case the whole expression (\ref{sing}) should vanish because it is
an odd matrix. This shows, in fact, that non-degeneracy of
$D_{\s}$ would allow one to remove the branch cut singularity by
means of a regular gauge transformation. In our present analysis
we find, however, that the traceless part of the matrix $D_{\s}$
is highly degenerate, it has four $+\kappa$ and four $-\kappa$
eigenvalues. Therefore, expression (\ref{sing}) and, as a
consequence, the whole monodromy cannot be
diagonalized\footnote{This was also explicitly verified by
computing ${\rm g}_0^{-1}\L_{\s}^{(1)}{\rm g_0}$ with $\gg_0$
given by eq.(\ref{ge}).} around the singular point by means of a
regular gauge transformation. Let us also note that due to
degeneracy of $D_{\s}$ the matrix $\gg_0$ is fixed only up to
multiplication from the left by any supermatrix which commutes
with $D_{\s}$. This freedom  is still not enough to make
(\ref{sing}) to vanish. Indeed, would it be the case for some
$\gg_0$ and $\gg_1$ then we would perform the same asymptotic
analysis as before but for the new Lax connection  and  find that
$I$ in eq.(\ref{wbc}) is conserved which is not the case!
 Degeneracy of $D_{\s}$
at a singular point is welcome, otherwise we could remove this
singularity by means of a regular gauge transformation which would
mean only a fake presence of fermionic degrees of freedom in the
theory.  The absence of the local conservation laws in the leading
asymptotic expansion of the Lax connection around singular point
is therefore related to the fact that the connection is not
diagonalizable at this point by a regular element.

\bigskip

Let us further note that in the absence of fermions  the full
$\AdS$ model has the following integrals of motion \bea
\label{Ipm} I_{\pm}=\int_0^{2\pi}\frac{{\rm
d\s}}{2\pi}\sqrt{(\P^M\pm \acute{u}^M)^2} \, . \eea Indeed, we
have \bea \nonumber \pa_{\tau}I_{\pm}=\int_0^{2\pi}\frac{{\rm
d\s}}{2\pi}\frac{(\P^M\pm \acute{u}^M)}{\sqrt{(\P\pm
\acute{u})^2}}\Big(-\frac{e^{2\phi}}{p^+}u^M\P^2+\frac{1}{p^+}\pa_{\s}(e^{2\phi}\acute{u}^M)
\pm\pa_{\s}\Big(\frac{e^{2\phi}}{p^+}\P^M\Big) \Big)\, ,\eea where
we have used the equations (\ref{sbf}). Using the constraints
$\P^Mu^M=u^M\acute{u}^M=0$ it is not difficult to see that \bea
\pa_{\tau}I_{\pm}=\pm\int_0^{2\pi }\frac{{\rm d\s}}{2\pi p^+
}\pa_{\s}\Big(e^{2\phi}\sqrt{(\P\pm \acute{u})^2}\Big)=0\, . \eea
If we set   the momentum $\P^M$ to zero in eq.(\ref{Ipm}) the
integral becomes just a
 length of the string ``drawn'' on a
five-sphere. When
 the string moves in time the length itself is not a
conserved quantity.

Finally, we remark that the integral $I_-$
arises upon the expansion of the Lax connection around $\lambda=1$
while $I_+$ emerges from the expansion near
 $\lambda=-1$. In the appendix C we shall
present an independent derivation of $I_{\pm}$ for the bosonic
string model.

%\begin{equation}
%g={\scriptsize \left(
%\begin{array}{cccccccc}
%  0 & 0 & 0 & 0 & -\frac{1}{i\kappa}(iz^1+z^4) & -\frac{1}{i\kappa}(z^2+iz^3) & 0 & 1 \\
%  0 & 0 & 0 & 0 & \frac{1}{i\kappa}(z^2-iz^3) & \frac{1}{i\kappa}(iz^1-z^4) & 1 & 0 \\
%  0 & \frac{e^{i\phi}}{\sqrt{2}pp}(i\kappa+\P_{\phi}-\acute{\phi}) & 0 & 1 & 0 & 0 & 0 & 0 \\
%  -\frac{i\sqrt{2}e^{-\phi}pp}{i\kappa-\P_{\phi}+\acute{\phi}} & 0 & 1 & 0 &  0 & 0 & 0 & 0 \\
%  0 & 0 & 0 & 0 & \frac{1}{i\kappa}(iz^1+z^4) & \frac{1}{i\kappa}(z^2+iz^3) & 0 & 1 \\
%  0 & 0 & 0 & 0 & \frac{1}{i\kappa}(iz^3-z^2) & \frac{1}{i\kappa}(-iz^1+z^4) & 1 & 0 \\
%  0 &  -\frac{ie^{\phi}}{\sqrt{2}pp}(i\kappa-\P_{\phi}+\acute{\phi}) & 0 & 1 & 0 & 0 & 0 & 0 \\
%  \frac{i\sqrt{2}e^{-\phi}pp}{i\kappa+\P_{\phi}-\acute{\phi}} & 0 & 1 & 0 & 0 & 0 & 0 & 0
%\end{array}\right)
%},
%\end{equation}

\section{Integrability  in   ${\rm AdS}_3\times {\rm S}^1$ sector}
%%%%%%%%%%%%%%%%%%%%%%%%%%%%%%%%%%%%%%%%%%%%%%%%

In this section we will study the integrability
 properties of the Lax
connection in greater detail by specifying to the
%supersymmetric
${\rm AdS}_3 \times {\rm S}^1$ subsector
\rf{p},\rf{g}.\foot{String states associated to semiclassical
solutions in this  sector may be related to the closed
$\su(1,1|1)$ sector on the gauge theory side \ci{beis}. However
this needs further investigation.} Restricting to this subsector
 will allow us to reduce the
number of dynamical variables in a consistent way while preserving
the nontrivial
 features of the superstring sigma model.
 The non-vanishing bosonic fields are then $\phi$ and
$u_2$, $u_3$ (with $u_2^2+u_3^2=1$) plus their conjugate momenta
$\P_{\phi}$, $\P_{2},\P_3$. The fermionic degrees of freedom are
$\theta_1$, $\theta^1$ and $\eta_4,\eta^4$. Upon this reduction
the original string equations are dramatically simplified, and we
present them in Appendix B in terms of new variables.

In section 2 we showed  that the leading asymptotics of the Lax
connection around regular points $\lambda\to 0$ and
$\lambda\to\infty$ always reproduce (up to rotations by constant
matrices) the non-abelian Noether charge of the global
$\psu(2,2|4)$ symmetry. Again, in terms of  matrix $\su(2,2|4)$
representation, we should expect that only its traceless part is
conserved. The general form of the Noether current is given by
eq.(\ref{cc}). Putting this current in our explicit matrix
representation we found the following equation for its divergence:
\bea
\pa_{\tau}\underbrace{\Big(-p^+e^{-2\phi}a_{\tau}^{(2)}+\frac{1}{2}q'_{\s}\Big)}_{\tau-{\rm
component }}+
\pa_{\sigma}\underbrace{\Big(\frac{e^{2\phi}}{p^+}a_{\s}^{(2)}-\frac{1}{2}q'_{\tau}\Big)}_{\s-{\rm
component }} =i\Lambda \mathbb{I}_{8\times 8}\, .
 \eea
Here  the diagonal $\Lambda$-term appears to be \bea \nonumber
\Lambda=-\frac{p^+}{4}
\pa_{\tau}\Big(3\eta_4\eta^4+\theta_1\theta^1\Big)
-\frac{i}{4}\pa_{\s}\Big(e^{2\phi}\eta^4\theta^1(u_2-iu_3)+e^{2\phi}\eta_4\theta_1(u_2+iu_3)\Big)\,
.
 \eea
As expected, the traceless part  is perfectly conserved.

Let us now assume that just as in the bosonic case
the classical
solutions we consider carry only the Cartan (diagonal) charges of the
$\psu(2,2|4)$ algebra. Utilizing our $8\times 8$ matrix
representation,
 we find after some tedious computation  that the
traceless diagonal part of the conserved charge \bea {\mathbf
Q}=\int {\rm d}\s J^{\tau}=\frac{1}{2}{\rm diag}(p_1,\ldots,
p_8)\, \eea can be presented in the following way
\begin{center}
\begin{tabular}{ll}
\multicolumn{1}{c}{{\small AdS}~~~~~~} &
\multicolumn{1}{c}{{\small Sphere}} \\
$p_1=\dD-J^{+-}-J^{x\bar{x}}$,~~~~~~~~ &
$p_5=2i J^1_{~1}$, \\
$p_2=\dD+J^{+-}+J^{x\bar{x}}$,~~~~~~~~ &
$p_6=2i J^2_{~2}$,\\
$p_3=-\dD+J^{+-}-J^{x\bar{x}}$,~~~~~~~~ &
$p_7=2i J^3_{~3}$,\\
$p_4= -\dD-J^{+-}+J^{x\bar{x}}$,~~~~~~~~ & $p_8=2i J^4_{~4}$\, .
\end{tabular}
\end{center}
\vskip 0.3cm \noindent Here the generators
$\dD,J^{+-},J^{x\bar{x}}$ and $J^i_{~j}$ have the following
explicit form \bea \dD&=&\int {\rm d}\s
(x^+\P^-+x^- p^+-\P_{\phi}) \\
J^{+-}&=&\int{\rm d}\s  (  x^+\P^--x^-p^+) \\
J^{x\bar{x}}&=&\int {\rm d}\s  ( -\frac{i}{2}p^+
\theta^2+\frac{i}{2}p^+\eta^2) \,
\\
\label{su4} J^i_{~j}&=&\int{\rm d}\s  \big[
\frac{i}{2}(\rho^{MN})^{i}{}_{j}u^M\P^N+p^+\theta^i\theta_j+p^+\eta^i\eta_j
-\frac{1}{4}\delta^i_j p^+(\theta^2+\eta^2) \big]\, . \eea The reader
can now recognize that these integrals are precisely the
kinematical generators of the light-cone superstring
\cite{Metsaev:2000yu} specified for our reduced model. Here $\dD$ is
the generator of scale transformations, $J^{+-}$ and $J^{x\bar{x}}$
generate rotations in the $(x^{+},x^{-})$ and $(x,\bar{x})$ planes
respectively, and $J^i_{~j}$ are the generators of $\su(4)$. All
these generators have non-negative charge w.r.t. to $J^{+-}$. We
have written down the $\su(4)$ generators for the general case but
note that for our ${\rm AdS_3}\times {\rm S}^1$ model all its {\it
non-diagonal} components, $J^i_{~j}$ with $i\neq j$, vanish.

\medskip
It is useful to compare the leading asymptotics of the Lax
connection (the monodromy matrix) we just obtained for the reduced
model with that of the bosonic $\AdS$ sigma-model. The latter was
recently obtained in \cite{Arutyunov:2004yx} (see also
\cite{bksa}) by using another (uniform) gauge choice. In this
gauge the Hamiltonian ${\rm H}$ coincides with the energy of the
string defined with respect to
 the global AdS time.  The leading asymptotics of the
Lax connection on the Cartan solutions are  found to be
\cite{Arutyunov:2004yx}
\begin{center}
\begin{tabular}{ll}
\multicolumn{1}{c}{{\small AdS}~~~~~~} &
\multicolumn{1}{c}{{\small Sphere}} \\
$p_1\sim {\rm H}-S_1-S_2$ ,~~~~~~ &
$p_5\sim -J_1-J_2+J_3$, \\
$p_2\sim {\rm H}+S_1+S_2$,~~~~~~ &
$p_6\sim -J_1+J_2-J_3$,\\
$p_3\sim -{\rm H}+S_1-S_2$,~~~~~~ &
$p_7\sim J_1-J_2-J_3$,\\
$p_4\sim -{\rm H}-S_1+S_2$,~~~~~~ & $p_8\sim J_1+J_2+J_3$\, .
\end{tabular}
\end{center}
Here $S_1$ and $S_2$ are Cartan generators of the unbroken so(4)
symmetry (AdS spins), and $(J_1,J_2,J_3)$ are the Cartan
components of the so(6) angular momentum. We see that the
asymptotics of the Lax connection around $\lambda=0$ and
$\lambda=\infty$  are {\it the same} (up to unessential numerical
prefactors and permutations of $p$'s),    provided we make an
obvious identification ${\rm H}=\dD$, $J^{+-}=S^1$,
$J^{x\bar{x}}=S_2$ and properly relate the so(6) labels with
$J^i_{~i}$ of $\su(4)$. In both the bosonic and fermionic cases
the diagonal components of the Lax connection are expressed in
terms of the Cartan charges in the same way. It is  very
suggestive that if we could repeat our computation for the full
$\AdS$
 superstring model
(which seems however a difficult task due to the
large number of
fields) we would be able  to represent again
the result in terms of
the kinematical generators,
 and in terms of these generators it
would look the same as for the reduced ${\rm AdS}_3\times {\rm
S}^1$ model.

\medskip
\medskip
\medskip

We would like also to understand if and how eq.(\ref{basic}) can
still be used to obtain a non-trivial information about
integrability properties of the system. By trial and error we
found that the decomposition (\ref{factor}) in terms of
derivatives of local quantities can be achieved for the ${\rm
AdS}_3 \times {\rm S}^1$ sector.  As the result,  we obtained the
following conserved integral
\begin{equation}
\label{sIm} I_-^{\rm f}=\int_0^{2\pi}\frac{{\rm
d\s}}{2\pi}\Big(\sqrt{(\P^M-\acute{u}^M)^2} + p^+ \eta_4
\eta^4\Big)\, , ~~~~~~~~M=2,3\, .
\end{equation}
This integral has the required bosonic limit and therefore
can be viewed as a supersymmetrization of the bosonic integral
$I_-$ in \rf{Ipm}. Quite remarkably, in our present case
 (and also for
the {\it bosonic} string in ${\rm AdS}_5\times {\rm S}^1$) the
constraints on $\P^M$ and $u^M$ are so powerful that they force
the expression under the square root in eq.(\ref{sIm}) to becomes
a perfect square\footnote{The same phenomenon was observed for the
bosonic string on ${\rm AdS}_5\times {\rm S}^1$ treated in the
uniform gauge \cite{Arutyunov:2004yx}. } \bea
(\P_M-\acute{u}_M)^2=\frac{1}{u_3^2}(\P_2-\acute{u}_2)^2 \, .\eea
Since $\acute{u_2}/u_3=\acute{u_2}/(1+u_2^2)$ is a total
derivative it can be omitted and we end up with \bea I_-^{\rm
f}=\int_0^{2\pi}\frac{{\rm d\s}}{2\pi}\Big(\frac{\P_2}{u_3} + p^+
\eta_4 \eta^4\Big)\, . \eea To understand the meaning of this
integral we recall eq.(\ref{su4}) for the Noether charge
corresponding to the $\su(4)$ symmetry.
 As we have already pointed out, the non-diagonal
components of $J^i_{~j}$ vanish. Taking into account that
$u_3\P_2-u_2\P_3=\P_2/u_3$ for the diagonal components we obtain
\bea \nonumber J^1_{~1}&=&\int  {\rm d}\s (
-\frac{\P_2}{2u_3}-\frac{3}{4}p^+\theta_1\theta^1+\frac{1}{4}p^+\eta_4\eta^4
)\, ,
\\
\nonumber J^2_{~2}&=&J^3_{~3}=\int {\rm d}\s (
\frac{\P_2}{2u_3}+\frac{1}{4}p^+\theta_1\theta^1+\frac{1}{4}p^+\eta_4\eta^4)\,
 ,
\\
\nonumber J^4_{~4}&=&\int {\rm d}\s (
-\frac{\P_2}{2u_3}+\frac{1}{4}p^+\theta_1\theta^1-\frac{3}{4}p^+\eta_4\eta^4
) \, . \eea The Dynkin labels $[a_1,a_2,a_3]$ of an  $\su(4)$
representation are related to the Cartan components as
$$a_1\sim
J^1_{~1}-J^4_{~4}\ ,  \ \ a_2\sim J^4_{~4}-J^3_{~3}\ ,
   \ \   a_3\sim J^1_{~1}+J^4_{~4}+2J^3_{~3}\ . $$
Substituting here the expressions for $J^i_{~i}$ we find
\bea
a_1\sim p^+\int {\rm d}\sigma (\eta_4\eta^4-\theta_1\theta^1)\,
,~~~~ a_2\sim -\int {\rm d}\sigma (\P_2/u_3+p^+\eta_4\eta^4)\,
 \eea
and $a_3=0$. We thus observe that the integral $I_{-}^{\rm f}$ is
proportional to  the Dynkin label $a_2$. Similar consideration can
be repeated for $\lambda\to -1$.

\bigskip

To summarize, we have shown that the leading asymptotics of the
monodromy matrix around the branch cut singularity at $\lambda=1$
is related to one of the global charges of the model which is
proportional to the central Dynkin label of the corresponding
$\su(4)$ irrep. This asymptotic behavior in the superstring theory
reminds the corresponding bosonic string pattern
\cite{Ka,Arutyunov:2004yx}.

\medskip

\section{Concluding remarks }

%%%%%%%%%%%%%%%%%%%%%%%%%%%%%%%%%%%%%%%%%%%
In this paper we have obtained the Lax representation for the
Hamiltonian of the classical superstring theory on $\AdS$ in the
light-cone gauge. The Lax connection depends on physical degrees
of freedom only and it is explicitly realized in terms of
$\su(2,2|4)$ matrices.
%The lack of a covariant
%Hamiltonian treatment for quantum superstring renders important This is step forward towards clarifying integrable properties of the physical superstring which arises upon removing all gauge degrees of freedom.

We have found that in the presence of fermions a Lax pair does not
immediately imply the existence of {\it local} conservation laws.
It appears not possible to diagonalize the monodromy around a
singular point in the spectral parameter plane by a regular gauge
transformation. As a consequence, the r.h.s. of the conservation
equation (\ref{basic}) receives a non-trivial contribution from
fermionic fields. It is not clear a priory whether this
contribution can be written as a divergence of some {\it local}
current so that to be able to define a new improved current which
would be conserved. This makes the connection between the Lax pair
and the local integrals of motion not as straightforward as in the
purely bosonic case. Of course, around a generic point on the
spectral plane the monodromy matrix is diagonalizable and
generates (non-local) integrals of motion.

%This resembles what happens in quantum integrable systems where
%the notion of the Lax pair can be sometimes introduced but it does
%not guarantee the existence of integrals of motion. On the other
%hand, the account of fermionic degrees of freedom becomes {\it
%trully} relevant in the quantum theory, and, therefore, the
%fermionic obstruction to conservation laws might vanish as an
%expectation value between physical states, providing thereby
%conserved quantities on the physical subspace.
%This point deserves
%further investigation.

We observed that the superstring equations of motion can be
consistently truncated to supersymmetric subsectors which we
called ${\rm AdS}_3\times {\rm S}^{3}$ and ${\rm AdS}_3\times {\rm
S}^{1}$. Rather remarkably, in the latter case the fermionic
contribution to the conservation law for the eigenvalues of the
monodromy (around $\lambda =1$) can be represented as a divergence
of some local current. This allowed us to construct a local
integral of motion which includes  fermions and has the proper
bosonic limit.

Finally, we have proved quite generally that the leading term of
the asymptotics of the Lax connection around zero or infinity
gives the Noether charges of the global $\psu(2,2|4)$ symmetry.
For the case of the reduced ${\rm AdS}_3\times {\rm S}^{1}$ model
we have computed these charges explicitly in terms of the physical
fields. We then expressed the Cartan (diagonal) components of the
Noether charges in terms of the kinematical light-cone generators
finding the same relations as in the bosonic case. It is natural
to expect that such relations continue to hold also for the full
$\AdS$.

As for open problems, it would be desirable to find an efficient
way to generate the local integrals of motion from the Lax pair
for the general case. It would be also interesting to better
understand the meaning of the supersymmetric ${\rm AdS}_3\times
{\rm S}^{3}$ and ${\rm AdS}_3\times {\rm S}^{1}$ subsectors from
the dual gauge theory point of view.

\newpage
%%%%%%%%%%%%%%%%%%%%%%%%%%
\section*{Acknowledgments }
%%%%%%%%%%%%%%%%%%%%%%%%%%%

We are grateful to N. Beisert, S. Frolov,
 R. Metsaev, H. Nicolai, M. Staudacher, \mbox{B. Stefanski} and
K. Zarembo for useful discussions. The work of  G.~A. was
supported in part by the European Commission RTN programme
HPRN-CT-2000-00131 and by RFBI grant N02-01-00695. The  work of
A.A.T.  was supported  by the DOE grant DE-FG02-91ER40690  and
also by
 the INTAS contract 03-51-6346
and the RS Wolfson award.

While preparing this paper for submission we learned about the
interesting very recent paper \ci{bksz} which also investigates
the integrability of classical superstring theory on \mbox{$\AdS$}
but without fixing particular gauges.

\vskip 1cm
%\newpage

%%%%%%%%%%%%%%%%%%%%%%%%%%%%%%%%%%%%%%%%%%%%%%%%%%%%%%%%%%%%%%

\appendix
\section{Matrices}

Here we collect the information about various matrices we use
throughout the paper. We represent the generators of the
superconformal group by the $\su(2,2)$ matrices. In particular,
the generator of scaling transformations is chosen to be
\medskip
\begin{eqnarray}
{\rm D}={\scriptsize \frac{1}{2}\left(
\begin{array}{cccc}
  1 & 0 & 0 & 0 \\
  0 & 1 & 0 & 0 \\
   0 & 0 & -1 & 0 \\
   0 & 0 & 0 & -1
\end{array} \right)\, .}
\end{eqnarray}
The generators of translations are given by
\medskip
\begin{eqnarray}
\nonumber {\rm P}^0={\scriptsize \left(
\begin{array}{cccc}
  0 & 0 & 0 & 0 \\
  0 & 0 & 0 & 0 \\
   i & 0 & 0 & 0 \\
   0 & i & 0 & 0
\end{array} \right)},\hspace{0.3in}{\rm P}^1={\scriptsize \left(
\begin{array}{cccc}
  0 & 0 & 0 & 0 \\
  0 & 0 & 0 & 0 \\
   i & 0 & 0 & 0 \\
   0 & -i & 0 & 0
\end{array} \right)},\hspace{0.3in}
{\rm P}^2={\scriptsize \left(
\begin{array}{cccc}
  0 & 0 & 0 & 0 \\
  0 & 0 & 0 & 0 \\
   0 & i & 0 & 0 \\
   i & 0 & 0 & 0
\end{array} \right)},\hspace{0.3in}{\rm P}^3={\scriptsize \left(
\begin{array}{cccc}
  0 & 0 & 0 & 0 \\
  0 & 0 & 0 & 0 \\
   0 & 1 & 0 & 0 \\
   -1 & 0 & 0 & 0
\end{array} \right)}
\end{eqnarray}
The conformal boosts are defined as \bea {\rm K}^i=({\rm
P}^i)^t,~~~~\mbox{for}~~i=0,3;\hspace{0.3in}{\rm K}^i=-({\rm
P}^i)^t,~~~~\mbox{for}~~i=1,2 \, .\eea

\medskip

The generators of $\su(4)$ can be given in terms of the following
${\rm so}(5)$-gamma matrices

\begin{eqnarray}
\nonumber &&\Gamma^1={\scriptsize \left(
\begin{array}{cccc}
  0 & 0 & 0 & -1 \\
  0 & 0 & 1 & 0 \\
   0 & 1 & 0 & 0 \\
   -1 & 0 & 0 & 0
\end{array} \right)},\hspace{0.3in}\Gamma^2={\scriptsize \left(
\begin{array}{cccc}
  0 & 0 & -i & 0 \\
  0 & 0 & 0 & i \\
   i & 0 & 0 & 0 \\
   0 & -i & 0 & 0
\end{array} \right)},\hspace{0.3in}\Gamma^3={\scriptsize \left(
\begin{array}{cccc}
  0 & 0 & -1 & 0 \\
  0 & 0 & 0 & -1 \\
   -1 & 0 & 0 & 0 \\
   0 & -1 & 0 & 0
\end{array} \right),}\\
&&\Gamma^4={\scriptsize \left(
\begin{array}{cccc}
  0 & 0 & 0 & i \\
  0 & 0 & i & 0 \\
   0 & -i & 0 & 0 \\
   -i & 0 & 0 & 0
\end{array} \right)},\hspace{0.3in}\Gamma^5={\scriptsize \left(
\begin{array}{cccc}
  1 & 0 & 0 & 0 \\
  0 & 1 & 0 & 0 \\
   0 & 0 & -1 & 0 \\
   0 & 0 & 0 & -1
\end{array} \right)}\, .
\end{eqnarray}
Notice that these generators have charge $-1$ with respect to $J$,
{\it i.e.} $ J(\Gamma^A)^t J =-\Gamma^A$. The so(6) generators are
spanned by $i \Gamma^A$ and $[\Gamma^A,\Gamma^B]$.

\medskip
To present the $\kappa$-fixed light-cone string equations of
motion (\ref{abf}), (\ref{sbf}) and (\ref{fermion}) we need the
matrices $\rho_{ij}^M$. These matrices are used to construct the
so(6) $\gamma$-matrices $\gamma^M$ in the chiral representation
\begin{equation}
\gamma^M=\left( \begin{array}{cc}
  0 & (\rho^M)^{ij} \\
  \rho_{ij}^M & 0
\end{array} \right)\, .
\end{equation}
they satisfy the following algebra
\begin{eqnarray}
(\rho^M)^{il} \rho_{lj}^N+(\rho^N)^{il} \rho_{lj}^M = 2 \delta^{M
N} \delta^i_j
\end{eqnarray}
and the completeness condition \bea \rho_{ij}^M
(\rho^M)^{kl}=2(\delta_i^l \delta_j^k -\delta_i^k \delta_j^l) \, .
\eea They also have the following symmetry properties \bea
\rho_{ij}^M=-\rho_{ji}^M,\hspace{0.3in}(\rho^M)^{ij} =
-(\rho_{ij}^M)^*\, , ~~~~~
\rho_{ij}^M=\frac{1}{2}\epsilon_{ijkl}(\rho^M)^{kl}\, .  \eea In
this  paper we made use of the following explicit form
\begin{eqnarray}
\nonumber \rho^1={\scriptsize\left(
\begin{array}{cccc}
  0 & 0 & i & 0 \\
  0 & 0 & 0 & i \\
   -i & 0 & 0 & 0 \\
   0 & -i & 0 & 0
\end{array} \right)},\hspace{0.3in}\rho^2={\scriptsize\left(
\begin{array}{cccc}
  0 & 0 & 0 & 1 \\
  0 & 0 & 1 & 0 \\
   0 & -1 & 0 & 0 \\
   -1 & 0 & 0 & 0
\end{array} \right)},\hspace{0.3in}\rho^3={\scriptsize \left(
\begin{array}{cccc}
  0 & 0 & 0 & -i \\
  0 & 0 & i & 0 \\
   0 & -i & 0 & 0 \\
   i & 0 & 0 & 0
\end{array} \right)},\\
\rho^4={\scriptsize \left(
\begin{array}{cccc}
  0 & 0 & 1 & 0 \\
  0 & 0 & 0 & -1 \\
   -1 & 0 & 0 & 0 \\
   0 & 1 & 0 & 0
\end{array} \right)},\hspace{0.3in}\rho^5={\scriptsize \left(
\begin{array}{cccc}
  0 & i & 0 & 0 \\
  -i & 0 & 0 & 0 \\
   0 & 0 & 0 & -i \\
   0 & 0 & i & 0
\end{array} \right)},\hspace{0.3in}\rho^6={\scriptsize \left(
\begin{array}{cccc}
  0 & 1 & 0 & 0 \\
  -1 & 0 & 0 & 0 \\
   0 & 0 & 0 & 1 \\
   0 & 0 & -1 & 0
\end{array} \right)} \, .
\end{eqnarray}

\medskip
The matrices $\rho^{MN}$ are defined by
\begin{equation}
(\rho^{MN})^i_{\,j} \equiv \frac{1}{2} (\rho^M)^{il} \rho_{lj}^N -
\frac{1}{2} (\rho^N)^{il} \rho_{lj}^M \, .
\end{equation}
They satisfy the following completeness condition
\begin{equation}
(\rho^{MN})^i_{\,j}(\rho^{MN})^k_{\,l}=2\delta^i_j
\delta^k_l-8\delta^i_l \delta^k_j \, .
\end{equation}

\medskip

The matrix $\gg$ which diagonalizes $\L_{\a}^{(0)}=\gg
D_{\a}\gg^{-1}$ can be chosen to be the following even element
from ${\rm SU}(2,2|4)$: \bea \label{ge} \gg=\left(
\begin{array}{cc} {\rm h}_1 & 0 \\
                   0 & {\rm h}_2
\end{array}
\right)\, , \eea where \bea \nonumber {\rm h}_1={\footnotesize
\left(
\begin{array}{cccc}
0 & \frac{\sqrt{2}p^+e^{-\phi}}{q}  &  0   &   -\frac{\sqrt{2}p^+e^{-\phi}}{\bar{q}}\\
\frac{e^{\phi}\bar{q}}{\sqrt{2}p^+} &  0   &  -\frac{e^{\phi}q}{\sqrt{2}p^+} & 0 \\
0 & 1 & 0 & 1  \\
1 & 0 & 1 & 0
\end{array}\right)}
\, , ~~~~~ {\rm h}_2={\footnotesize \left(\begin{array}{cccc}
q^{14} & \bar{q}^{32} & -q^{14} & -\bar{q}^{32} \\
q^{32} & -\bar{q}^{14} & -q^{32} & \bar{q}^{14} \\
0 & 1 & 0 &  1 \\
1 & 0 & 1 & 0
\end{array}\right)
}\, .
\eea
%\begin{equation}
%\nonumber \gg={\footnotesize \left(
%\begin{array}{cccccccc}
%  0 & 0 & 0 &  -\frac{\sqrt{2}p^+e^{-\phi}}{\bar{q}} & 0 & 0 & 0 &  \frac{\sqrt{2}p^+e^{-\phi}}{q} \\
%  0 & 0 & -\frac{e^{\phi}q}{\sqrt{2}p^+} & 0 & 0 & 0 & \frac{e^{\phi}\bar{q}}{\sqrt{2}p^+} & 0 \\
%  0 & 0 & 0 & 1 & 0 & 0 & 0 & 1 \\
%  0 & 0 & 1 & 0 &  0 & 0 & 1 & 0 \\
%  -q^{14} & -\bar{q}^{32} & 0 & 0 & q^{14} & \bar{q}^{32} & 0 & 0 \\
%  -q^{32} & \bar{q}^{14} & 0 & 0 & q^{32} & -\bar{q}^{14} & 0 & 0 \\
%  0 &  1 & 0 & 0 & 0 & 1 & 0 & 0 \\
%  1 & 0 & 0 & 0 & 1 & 0 & 0 & 0
%\end{array}\right)
%}\, .
%\end{equation}
Here we used the concise notation \bea \nonumber
q^{14}&=&\frac{1}{\kappa}(\P^{1}-\acute{u}^1-i(\P^{4}-\acute{u}^4))\, , \\
\nonumber
q^{32}&=&\frac{1}{\kappa}(\P^{3}-\acute{u}^3-i(\P^{2}-\acute{u}^2))\, , \\
\nonumber
 q&=&\kappa-i(\P_{\phi}-\acute{\phi})\,
 \eea
 where $\kappa$ is given by eq.(\ref{kappa})
and $\bar{q}$ denotes the corresponding complex conjugate.

\section{String equations
 in ${\rm AdS}_3\times {\rm S}^1$ sector}
To discuss the ${\rm AdS}_3 \times {\rm S}^1$ truncation it is
convenient to solve explicitly the constrains on $u_2,u_3,{\P}_2$
and ${\P}_3$. To this end we parametrize these fields in terms of
two real variables $u$ and $\P_u$ as follows
\begin{eqnarray}
u_2+i u_3=e^{i u},\hspace{0.2in}u_2-i u_3=e^{-i u}\\
{\P}_2 = {\P}_u u_3,\hspace{0.2in}{\P}_3=-{\P}_u u_2\, .
\end{eqnarray}
We also define $\eta\equiv\eta_4$, $\bar{\eta}\equiv\eta^4$ and
$\theta\equiv\theta_1$, $\bar{\theta}\equiv \theta^1$. Then the
Hamiltonian simplifies to
\begin{eqnarray}
{\cal H}=\frac{e^{2 \phi}}{2p^+}\Big({\P}_\phi^2
+{\P}_u^2+\acute{u}^2+\acute{\phi}^2-2p^+ \eta\bar{\eta} {\P}_u
-(2 e^{i u} p^+ \eta \acute{\theta}+~~{\rm c.c.})\Big)\, ,
\end{eqnarray}
while the equations of motion take the form
\begin{eqnarray}
\nonumber
&&\dot{\phi}=\frac{e^{2\phi}}{p^+}\P_\phi\, , \\
\nonumber
&&\dot{\P}_\phi=-\frac{e^{2\phi}}{p^+}\Big(\P_\phi^2+\P_u^2+\acute{u}^2-\phi''-\phi'^2-2p^+
\eta \bar{\eta} -(2p^+ e^{-i u} \bar{\eta} \acute{\bar{\theta}}+~{\rm c.c})\Big)\, ,\\
\nonumber &&\dot{u}=\frac{e^{2\phi}}{p^+}(p^+ \eta
\bar{\eta}-\P_u)\, ,
\hspace{0.2in}\dot{\P}_u=-\frac{e^{2\phi}}{p^+}\Big(2\acute{u}\acute{\phi}+u''+(i
p^+ e^{i u} \eta \theta +~{\rm c.c})\Big)\, ,\\
&&\dot{\eta} =-i\frac{e^{2\phi}}{p^+}(\P_u \eta
+e^{-iu}\acute{\bar{\theta}}),
\hspace{0.3in}\dot{\theta}=-\frac{i}{p^+}\partial_\sigma(e^{2\phi-i
u} \bar{\eta})\, .
\end{eqnarray}

\section{Integrability of bosonic strings in $\AdS$}
Here we describe yet another method to obtain the
leading asymptotics of the Lax connection around poles at
$\lambda=\pm 1$. This discussion uses  the general method
developed in \cite{Arutyunov:2004yx}.

Following \cite{Arutyunov:2004yx} we introduce the
matrix $g$
\begin{equation}
g=\left( \begin{array}{cc}
  g_a & 0 \\
  0 & g_s
\end{array} \right)\, ,
\end{equation}
where $g_a$ and $g_s$ are the following $4 \times 4$ matrices
\begin{equation}
g_a={\scriptsize \left( \begin{array}{cccc}
  0 & {\cal Z}_3 & -{\cal Z}_2 & {\cal Z}_1^* \\
  -{\cal Z}_3 & 0 & {\cal Z}_1 & {\cal Z}_2^* \\
   {\cal Z}_2 & -{\cal Z}_1 & 0 & -{\cal Z}_3^* \\
    -{\cal Z}_1^* & -{\cal Z}_2^* & {\cal Z}_3^* & 0
\end{array} \right)}, \hspace{0.3in}
g_s={\scriptsize \left( \begin{array}{cccc}
  0 & {\cal Y}_1 & -{\cal Y}_2 & {\cal Y}_3^* \\
  -{\cal Y}_1 & 0 & {\cal Y}_3 & {\cal Y}_2^* \\
   {\cal Y}_2 & -{\cal Y}_3 & 0 & {\cal Y}_1^* \\
    -{\cal Y}_3^* & -{\cal Y}_2^* & -{\cal Y}_1^* & 0
\end{array} \right) }\, .
\end{equation}
Here the complex embedding coordinates ${\cal Z}_k$ for the ${\rm
AdS}_5$ space and ${\cal Y}_k$ for the five-sphere are \bea
\nonumber {\cal Z}_1 &=& Z_1+ i Z_2\, , ~~~~~{\cal Z}_2 = Z_3+ i
Z_4\, ,
~~~~~{\cal Z}_3 = Z_0+ i Z_5 \\
\nonumber {\cal Y}_1&=&u^1+i u^2\, , ~~~~~~{\cal Y}_2=u^3+i u^4\,
,~~~~~~~{\cal Y}_3=u^5+i u^6 \, ,\eea where $u^Mu^M=1$ and
\begin{equation}
\nonumber -Z_0^2+Z_1^2+Z_2^2+Z_3^2+Z_4^2-Z_5^2=-1 \, .
\end{equation}
The variables $Z_A$ can be  expressed in terms of the coordinates
parametrizing the light-cone equations of motion  as follows
\begin{eqnarray}
\nonumber Z_0&=&\frac{1}{2}\Big(e^\phi+2(x \bar{x}+x^+
x^-)e^{\phi}+e^{-\phi}\Big),\hspace{0.3in}Z_5=\frac{e^{\phi}}{\sqrt{2}}(x^+-x^-),\\
\nonumber Z_1&=&\frac{1}{2}\Big(e^\phi-2(x \bar{x}+x^+
x^-)e^\phi-e^{-\phi}\Big),\hspace{0.3in}
Z_2=\frac{e^{\phi}}{\sqrt{2}}(x^+
+x^-),\\
\nonumber
Z_3&=&\frac{e^{\phi}}{\sqrt{2}}(x+\bar{x}),\hspace{0.3in}Z_4=-i\frac{e^{\phi}}{\sqrt{2}}(x-\bar{x}).
\end{eqnarray}
Introducing the currents
\begin{equation}
A_\alpha=(\partial_\alpha g) g^{-1}
\end{equation}
one can check that the equations of motion (\ref{abf}) and
(\ref{sbf}) with all fermions switched off can be written in the
form
\begin{equation}
\partial_\alpha(\gamma^{\alpha \beta} A_\beta) = 0 \, ,
\end{equation}
where $\gamma^{\a\b}$ is the Weyl-invariant combination of the 2d
metric (\ref{metric}).

Defining the following projectors
\begin{equation}
A_\alpha^{\pm}=(P^{\pm})_\alpha^\beta A_\beta,\hspace{0.3in}
(P^{\pm})_\alpha^\beta=\delta_\alpha^\beta \mp \gamma_{\alpha
\rho} \epsilon^{\rho \beta}
\end{equation}
we construct the projections of the current $A_{\a}$:
\begin{equation}
A_\tau^{\pm}=A_\tau \pm \frac{1}{p^+}e^{2\phi}A_\sigma,
\hspace{0.3in}A_\sigma^{\pm}=A_\sigma \pm p^+ e^{-2\phi}A_\tau \,
.
\end{equation}
We then use them to construct the Lax connection with a spectral
parameter $\lambda$
\begin{equation}
L_\alpha=\frac{A_\alpha^+}{2(1-\lambda)}+\frac{A_\alpha^-}{2(1+\lambda)}
\, .
\end{equation}
The string equations of motion imply that this connection has zero
curvature
\begin{equation}
\partial_\tau L_\sigma -\partial_\sigma L_\tau-[ L_\tau,L_\sigma]
=0\, .
\end{equation}
This way  we obtain a Lax operator for the bosonic Hamiltonian
(\ref{H}).

 As was discussed in section 4 we can also obtain the local
integrals of motion by expanding the Lax operator $L_\sigma$
around the poles ${\lambda}=\pm1$ and further diagonalize it.
Rather remarkably, at leading order in $1/(1\mp \lambda)$ we
obtain two identical blocks for the ${\rm AdS}_5$ and ${\rm S}^5$
parts:
\begin{equation}
L_\sigma \rightarrow i\left(
\begin{array}{cccc}
  \mathscr{I}_{\pm} & 0 & 0 & 0 \\
  0 &  \mathscr{I}_{\pm} & 0 & 0 \\
   0 & 0 & - \mathscr{I}_{\pm} & 0 \\
   0 & 0 & 0 & - \mathscr{I}_{\pm}
\end{array} \right)\, ,
\end{equation}
where $$ \mathscr{I}_{\pm}=\sqrt{({\cal P}^M \pm \acute{u}^M)({\cal
P}^M \pm \acute{u}^M)}.$$ According to the general theory, then
\begin{equation}
I_{\pm}=\int_0^{2\pi}\frac{{\rm d}\sigma}{2\pi}\sqrt{({\cal P}^M
\pm \acute{u}^M)^2}\,
\end{equation}
are local integrals of motion.

%\def \bi {bibitem}

%%%%%%%%%%%%%%%%%%%%%%%%%%%%%%%%%%%%%%%%%%%%%%%%%

\end{document}